\documentclass[twocolumn]{aastex63}

\usepackage{soulutf8} 
\usepackage{hyperref} 
\usepackage[utf8]{inputenc} 
\usepackage{enumerate}
\usepackage{color}
\definecolor{wildwatermelon}{rgb}{0.99, 0.42, 0.52}
\definecolor{trolleygrey}{rgb}{0.5, 0.5, 0.5}
\definecolor{cobalt}{rgb}{0.0, 0.28, 0.67}
\definecolor{bleudefrance}{rgb}{0.19, 0.55, 0.91}
\definecolor{cerulean}{rgb}{0.0, 0.48, 0.65}

\def\heii{\ion{He}{2}~4686~\r{A} }
\def\hheii{\ion{He}{2}~4686~\r{A}}

\received{June 1, 2019}
\revised{January 10, 2019}
\accepted{\today}
\submitjournal{AJ}

\shorttitle{Exploratory Spectroscopy of mCV Candidates II}
\shortauthors{Oliveira et al.}
\graphicspath{{./}{figs/}}

\begin{document}

\title{Exploratory Spectroscopy of Magnetic Cataclysmic Variables Candidates \\
and Other Variable Objects II}

\correspondingauthor{A. S. Oliveira}
\email{alexandre@univap.br}

\author[0000-0001-6422-9486]{A. S. Oliveira}
\affil{IP\&D, Universidade do Vale do Paraíba, 12244-000,
    São José dos Campos, SP, Brazil}

\author[0000-0002-9459-043X]{C. V. Rodrigues}
\affiliation{Divisão de Astrofísica, Instituto Nacional de Pesquisas Espaciais, 12227-010, São José dos Campos, SP, Brazil}

\author{M. Martins}
\affil{IP\&D, Universidade do Vale do Paraíba, 12244-000,
    São José dos Campos, SP, Brazil}

\author[0000-0002-0396-8725]{M. S. Palhares}
\affil{IP\&D, Universidade do Vale do Paraíba, 12244-000,
    São José dos Campos, SP, Brazil} 

\author[0000-0003-1949-4621]{K. M. G. Silva}
\affiliation{Gemini Observatory, Casilla 603, La Serena, Chile}
\affiliation{European Southern Observatory,  Alonso de Córdova 3107, Vitacura, Santiago, Chile}

\author[0000-0001-6013-1772]{I. J. Lima }
\affil{Divisão de Astrofísica, Instituto Nacional de Pesquisas Espaciais, 12227-010, São José dos Campos, SP, Brazil}    
\affil{Department of Astronomy, University of Washington, Box 351580, Seattle, WA 98195, USA}

\author[0000-0002-0386-2306]{F. J. Jablonski}
\affiliation{Divisão de Astrofísica, Instituto Nacional de Pesquisas Espaciais, 12227-010, São José dos Campos, SP, Brazil}



\begin{abstract}

This is the second paper of a series presenting our search for magnetic Cataclysmic Variables (mCVs) among
candidates selected mostly from the Catalina Real-Time Transient Survey (CRTS).
We present the identification spectra, obtained at the SOAR Telescope, as well as magnitudes and Gaia distances for 45 objects. Of these, 39 objects are identified as CVs, from which 8 targets show observational characteristics of mCVs, being 7 polars and 1 intermediate polar. The remaining 31 CVs in our sample are probably non-magnetic systems, in low (22 systems) or high (9 systems) accretion states. Six targets of the sample are not CVs (5 AGNs and 1 T Tauri star). Among the 8 objects with mCV spectra, 6 are new classifications. Three polars were observed in low accretion state, either revealing photospheric features of the secondary star and allowing the estimation of their spectral type, or presenting H$\beta$ Zeeman components associated to the WD magnetic field. In addition to the results obtained in the first paper of the series, and depending on the confirmation of these classifications by observational follow-up, our results would increase the sample of known polars by about 9 percent. 

\end{abstract}

\keywords{binaries: close --- novae, cataclysmic variables --- stars: dwarf novae --- stars: variables: general --- techniques: spectroscopic}


\section{Introduction} \label{sec:intro}

Magnetic Cataclysmic Variables (mCVs) are relatively rare objects among the Cataclysmic Variables (a review on CVs may be found in \citealt{warner1995}). While the number of known CVs exceeds a thousand objects (e.g., the Ritter-Kolb catalogue \citep{2003A&A...404..301R} registers 1429 CVs in its 7.24 edition),
the number of confirmed mCVs to date amounts to about 120 polars and 60 intermediate polars, or IPs \citep{2003A&A...404..301R,2015SSRv..191..111F}. See \citet{1990SSRv...54..195C} and \citet{1994PASP..106..209P} for reviews on polars and on IPs, respectively. 

Most of the identified polars have been discovered by soft X-ray surveys, especially by the ROSAT satellite, while the majority of IPs, being harder X-ray sources, have been found by hard X-ray surveys like those performed by Swift and Integral telescopes \citep[e.g.,][]{2010MNRAS.401.2207S}. The discovery rate of CVs had a significant increase due to automated surveys such as the multicolor imaging and spectroscopy survey SDSS \citep{2000AJ....120.1579Y} and variability-based surveys like ASAS \citep{2002AcA....52..397P}, MASTER \citep{2010AdAst2010E..30L}, and CRTS \citep{2009ApJ...696..870D}. \citet{2011AJ....142..181S}, for instance, present 285 CVs confirmed by SDSS spectra, including 33 polars and 7 IPs. At the same time, \citet{2014MNRAS.443.3174B} report the discovery of more than a thousand CV candidates in the six first years of operation of CRTS, observing that polars are likely to be missed by the survey due to their high/low state behavior.

Many questions remain open in the understanding of CVs. Among these are questions about the evolution of mCVs and, in particular, the evolutionary link between the longer-period IPs and the shorter-period polars \citep{1984ApJ...285..252C,2004ApJ...614..349N}. 
Also, the strong magnetic field in polars probably affects the efficiency of the magnetic braking mechanism for angular momentum loss \citep{1994MNRAS.266L...1W}, changing the mCV evolutionary timescales. 
This has been recently confirmed by the first attempt to explain the orbital period distribution of polars using binary population synthesis  (Belloni et al., in preparation).
How the evolutionary scenario can take into account the existence of the recently discovered short-period IPs \citep{2014MNRAS.442.2580P}, and what is the exact role of the low accretion-rate Pre-Polars \citep{2015SSRv..191..111F} in this play? Furthermore, what leads polars to phases of reduced mass-transfer and low-brightness states?

Answers to these questions could emerge from enlarged samples of those scarce objects. 
Our primary intention in this work is to increase the number of known mCV systems. This paper continues the optical spectroscopic survey conducted to search for signatures of magnetic accretion initiated in \citet{2017AJ....153..144O} (hereafter Paper I). This survey consists in snapshot spectroscopy of a sample of objects selected mainly from the CRTS catalog. Frequently, the definite classification as a mCV is only possible with additional information, so some objects classified as mCVs in our survey should be targets of follow-up time-resolved observations to confirm their classification. 

In Paper I we found 32 CVs in a sample of 45 objects. Among those 32 CVs, we classified 22 objects as strong mCV candidates by their spectroscopic aspect, 13 of which being reported as mCV for the first time. This means that 70 percent of our initial sample in Paper I were CVs, while almost 50 percent of the same sample was classified as mCVs. The remaining 30 percent of the sample were classified as extragalactic sources or as other types of variable stars. The first results on follow-up studies of mCV candidates from Paper I are presented in \citet{2015MNRAS.451.4183S} and \citet{2019MNRAS.489.4032O}, confirming their polar classification.

In the present paper we describe the second part of our spectroscopic survey, composed of SOAR spectra of 44 objects selected from the CRTS catalog and 1 object reported as a polar candidate in the literature. Section~\ref{sec_selection} presents the method for the sample selection, while Section~\ref{sec_Gaia} brings a discussion on the distance measurements from Gaia DR2. Section \ref{sec_observations} describes the SOAR spectroscopic data. In Section~\ref{sec_classification_criteria}, we show the classification criteria. The spectroscopic results and the classification of each object are shown in Section \ref{sec_classification}. Section~\ref{sec_discussion} concludes the paper with a summary and a discussion of the main results.
 
\section{SAMPLE SELECTION}
\label{sec_selection}

Objects of this survey were selected by visual inspection of CRTS light curves of CV candidates, available in their website\footnote{\url{http://crts.caltech.edu}}. The exception is Larin~2, which was selected from The Astronomer's Telegrams because of its X-ray behavior and peculiar colors. In the light curves inspection, we searched for photometric characteristics of mCVs such as: 

\begin{enumerate}[i.]

\item Transitions between different levels of brightness of around 1~mag in objects without outbursts, typical of polar systems;

\item Intrinsic flux variability, which may indicate accretion processes. The flux dispersion should be larger than the observational error expected for the mean magnitude;

\item Few outbursts, in order to minimize the selection of dwarf novae (DN) systems, although we could not simply exclude objects with outbursts, since IPs can present such events.
 
\end{enumerate}

Mentions to possible mCV classification in the CRTS circulars were also considered in our selection. Besides, the objects should have suitable brightness and coordinates for SOAR observations.
No color criterion was applied to select objects for the survey. 
The RA ordered list of observed objects is shown in Table \ref{obs}. The identification of the objects from the CRTS is in the form TTTyymmdd:hhmmss$\pm$ddmmss, where TTT indicates one of the three dedicated telescopes of the survey -- CSS, SSS or MLS -- yymmdd is the discovery date of the transient and hhmmss$\pm$ddmmss are the target coordinates. In order to avoid confusion between transients discovered by the same telescope at the same date, and for simplicity, we will abbreviate these IDs as TTThhmm$\pm$dd. 

High-energy counterparts for all sources were searched, by their coordinates, in the High Energy Astrophysics Science Archive Research Center (HEASARC) database \footnote{\url{https://heasarc.gsfc.nasa.gov/}}, selecting all available X-ray missions. The counterpart candidates are mentioned in Section \ref{sec_classification} for each object we found a detection within each observatory position error circle.

\section{Gaia distances}
\label{sec_Gaia}

Table~\ref{obs} also presents the distances to our targets obtained by \citet{2018AJ....156...58B}, estimated from Gaia \citep{2016A&A...595A...1G, 2018A&A...616A...1G} Data Release 2 parallaxes. 
The inversion procedure (i.e. to determine accurate distance with realistic error bars from measured parallax) must account for the non-linearity of the inversion transformation and for the asymmetry of the probability distribution. It demands a Bayesian analysis and an assumption about the distances, known as prior \citep{2015PASP..127..994B}, and often leads to large error bars. The fractional parallax uncertainty ($f=\sigma_{\varpi}/\varpi$) is defined as the ratio of the ($1\sigma$) parallax uncertainty to the parallax measurement. Large fractional parallax uncertainties represent poor distance estimates, so discretion is required for $f\gtrsim 0.20$, which indeed is the case for  about 80 percent of the objects in the Gaia catalogue. Accordingly, the distance uncertainties presented in Table~\ref{obs} are expressed in terms of its 68 percent confidence interval, represented by distance lower and upper bounds. Negative parallax measurements and high $f$ ratio are usually associated to relatively distant and low S/N sources. Among the four transients which we classify as extragalactic sources in Sect.\ref{sec_extragal}, three have negative parallaxes, and for all of them the distance values are highly underestimated and incompatible with the extragalactic spectroscopic classification.

\section{Spectroscopic observations and data reduction}
\label{sec_observations}

All spectra were obtained in remote observing mode using the SOAR 4.1-m telescope at Cerro Pach\'on, Chile, with the Goodman High Throughput Spectrograph \citep{2004SPIE.5492..331C}. This instrument employs Volume Phase Holographic  (VPH) gratings to maximize throughput, reaching down to the atmospheric cutoff at 3200 \r{A}. It is equipped with a Fairchild 4096$\times$4096 CCD with $15\times15$ micron pixel$^{-2}$ (0.15$\arcsec$ pixel$^{-1}$), which we set to run in the $3\times3$ binning mode. The spectrograph was set to operate with the red camera, the 400 l mm$^{-1}$ grating, 1.0$\arcsec$ slit, and the GG~4555 blocking filter, yielding a spectral resolution of 6~\r{A} FWHM in the $4350-8350$~\r{A} spectral range.
Three exposures of each science target were obtained to remove cosmic rays, and the slit was aligned to the 
parallactic angle to avoid light losses due to the atmospheric differential refraction. Cu-He-Ar lamp exposures were obtained for wavelength calibration, which resulted in typical 0.6 \r{A}, or about 30~km~s$^{-1}$, r.m.s. residuals. This accuracy was assessed by the [\ion{O}{1}] 5577, 6300 and 6363~\r{A} telluric lines measurements. 
Telluric absorption features are also visible in the spectra at 6870, 7180 and 7630~\r{A}. Bias images and dome calibration flats were taken to correct for the detector read-out noise and sensitivity.  Spectra of spectrophotometric standards \citep{1992PASP..104..533H} were used for flux calibration. 
The data reduction, spectra extraction and wavelength, atmospheric extinction and flux calibrations were performed using standard IRAF\footnote{IRAF is distributed by the National Optical Astronomy Observatories, which are operated by the Association of Universities for Research in Astronomy, Inc., under cooperative agreement with the National Science Foundation.} routines.


\begin{longrotatetable}
\begin{deluxetable*}{llllcccclc}
\tabletypesize{\scriptsize}
\tablecaption{List of observed targets. 
\label{obs}}
\tablewidth{0pt}
\tablehead{
\colhead{Object name} & \colhead{Abbreviation} & \colhead{RA} & \colhead{Dec} & \colhead{Date obs.}  &
\colhead{Exp. time} & \colhead{Type\tablenotemark{\scriptsize{a}}} & \colhead{V} & \colhead{Distance} & \colhead{$f=\sigma_{\varpi}/\varpi$}\\
\colhead{} & \colhead{} & \colhead{(J2000)} & \colhead{(J2000)} & \colhead{}  &
\colhead{(s)} & \colhead{} & \colhead{(mag)} & \colhead{(pc)} & \colhead{}
}
\startdata                                                                                            
CSS101014:030535$-$092005  &  CSS0305$-$09 & 03:05:35 &$-$09:20:05  & 2018 Jan 16   &   600   &  D    &  19.4  & $ 1018 ^{+599 }_{-361} $  & 3.20\\
CSS081030:042434+001419    &  CSS0424+00   & 04:24:34 &+00:14:19    & 2018 Jan 16   &   480   &  D    &  17.9  & $ 909  ^{+178 }_{-130} $  & 0.16\\
CSS140903:044529+173745    &  CSS0445+17   & 04:45:29 &+17:37:45    & 2018 Jan 17   &   480   &  HA   &  18.3  & \nodata                   &\nodata\\
MLS101203:050253+171041    &  MLS0502+17   & 05:02:53 &+17:10:41    & 2018 Jan 17   &   480   &  HA   &  17.7  & $ 1699 ^{+853 }_{-482} $  & 0.48\\
MLS121018:050716+125314    &  MLS0507+12   & 05:07:16 &+12:53:14    & 2018 Feb 14   &   1200  &  HA   &  17.1  & $ 395  ^{+1269}_{-230} $  & 0.43\\
MLS101214:052959+184810    &  MLS0529+18   & 05:29:59 &+18:48:10    & 2018 Jan 16   &   180   &  HA   &  15.2  & $ 462  ^{+19  }_{-17 } $  & 0.04\\
CSS101108:054711$-$192525  &  CSS0547$-$19 & 05:47:11 &$-$19:25:25  & 2018 Jan 16   &   1200  &  P    &  18.3  & $ 868  ^{+474 }_{-241} $  & 0.32\\
SSS120320:061754$-$362654  &  SSS0617$-$36 & 06:17:54 &$-$36:26:54  & 2018 Jan 16   &   300   &  HA   &  13.6  & $ 910  ^{+198 }_{-140} $  & 0.17\\
SSS130329:064610$-$410416  &  SSS0646$-$41 & 06:46:10 &$-$41:04:16  & 2018 Jan 16   &   600   &  D    &  18.6  & $ 655  ^{+31  }_{-28 } $  & 0.05\\
MLS150118:072211+260255    &  MLS0722+26   & 07:22:11 &+26:02:55    & 2018 Feb 15   &   1200  &  D    &  19.6  & $ 902  ^{+765 }_{-335} $  & 0.50\\
CSS091111:073339+212201    &  CSS0733+21   & 07:33:39 &+21:22:01    & 2018 Feb 15   &   1200  &  D    &  20.3  & $ 1512 ^{+883 }_{-505} $  & 0.96\\
MLS110309:074223+172807    &  MLS0742+17   & 07:42:23 &+17:28:07    & 2018 Jan 17   &   1200  &  D    &  19.1  & $ 839  ^{+866 }_{-374} $  & 0.65\\
CSS100108:081031+002429    &  CSS0810+00   & 08:10:31 &+00:24:29    & 2018 Jan 16   &   1200  &  P    &  20.3  & $ 925  ^{+1000}_{-538} $  & 0.90\\
MLS170128:081210+040352    &  MLS0812+04   & 08:12:10 &+04:03:52    & 2018 Feb 14   &   2700  &  P    &  18.8  & $ 1356 ^{+543 }_{-324} $  & 0.33\\
CSS110114:091937$-$055519  &  CSS0919$-$05 & 09:19:37 &$-$05:55:19  & 2018 Jan 17   &   600   &  P    &  18.6  & \nodata                   &\nodata\\
SSS100505:093417$-$174421  &  SSS0934$-$17 & 09:34:17 &$-$17:44:21  & 2018 Jan 16   &   600   &  D    &  19.4  & $ 1447 ^{+431 }_{-283} $  & 0.26\\
MLS110303:095308+145837    &  MLS0953+14   & 09:53:08 &+14:58:37    & 2018 Jan 17   &   480   &  IP   &  19.0  & $ 448  ^{+61  }_{-48 } $  & 0.12\\
MLS120223:095322+094127    &  MLS0953+09   & 09:53:22 &+09:41:27    & 2018 Jan 18   &   480   &  E    &  20.4  & $ 604  ^{+617 }_{-263} $  & 0.62\\
SSS120111:095652$-$331216  &  SSS0956$-$33 & 09:56:52 &$-$33:12:16  & 2018 Jan 17   &   480   &  D    &  18.9  & $ 1487 ^{+1004}_{-519} $  & 0.65\\
SSS120215:102042$-$335002  &  SSS1020$-$33 & 10:20:42 &$-$33:50:02  & 2018 Jan 17   &   300   &  HA   &  15.4  & $ 1199 ^{+408 }_{-251} $  & 0.24\\
MLS110226:105051+102134    &  MLS1050+10   & 10:50:51 &+10:21:34    & 2018 Feb 14   &   1200  &  E    &  19.4  & $ 1495 ^{+750 }_{-502} $\tablenotemark{\scriptsize{b}}  & -0.58\\
MLS121203:113751+004218    &  MLS1137+00   & 11:37:51 &+00:42:18    & 2018 Feb 15   &   1800  &  E    &  20.1  & $ 1344 ^{+778 }_{-492} $\tablenotemark{\scriptsize{b}}  & -1.69\\
MLS150511:124539$-$073706  &  MLS1245$-$07 & 12:45:39 &$-$07:37:06  & 2018 Feb 14   &   480   &  D    &  18.8  & $ 1223 ^{+976 }_{-538} $  & 1.70\\
CSS120222:124602$-$202302  &  CSS1246$-$20 & 12:46:02 &$-$20:23:02  & 2018 Feb 14   &   480   &  D    &  20.4  & $ 437  ^{+522 }_{-158} $  & 0.31\\
MLS140518:124709$-$040758  &  MLS1247$-$04 & 12:47:09 &$-$04:07:58  & 2018 Apr 16   &   2700  &  E    &  20.5  & $ 1698 ^{+996 }_{-635} $\tablenotemark{\scriptsize{b}}  & -1.13\\
Larin2                     &  Larin2       & 12:48:51 &$-$41:26:54  & 2018 Apr16-17 &   300   &  T    &  18.7  & $ 125  ^{+2   }_{-2  } $  & 0.02\\
CSS140430:130136$-$052938  &  CSS1301$-$05 & 13:01:36 &$-$05:29:38  & 2018 Feb 14   &   480   &  D    &  18.7  & $ 1798 ^{+899 }_{-536} $  & 0.78\\
CSS120304:130755$-$202628  &  CSS1307$-$20 & 13:07:55 &$-$20:26:28  & 2018 Feb 14   &   480   &  D    &  19.0  & \nodata  & \nodata \\
CSS130419:132918$-$121622  &  CSS1329$-$12 & 13:29:18 &$-$12:16:22  & 2018 Feb 15   &   1200  &  E    &  19.5  & $ 1435 ^{+1214}_{-687} $  & 3.04\\
SSS130313:140204$-$365251  &  SSS1402$-$36 & 14:02:04 &$-$36:52:51  & 2018 Feb 15   &   480   &  D    &  20.4  & $ 1568 ^{+1581}_{-621} $  & 0.49 \\
SSS130312:140311$-$240135  &  SSS1403$-$24 & 14:03:11 &$-$24:01:35  & 2018 Apr 17   &   2700  &  D    &  21.8  & \nodata  &         \nodata  \\
MLS130513:141002$-$124809  &  MLS1410$-$12 & 14:10:02 &$-$12:48:09  & 2018 Apr 17   &   1200  &  HA   &  16.6  & $ 1357 ^{+1409}_{-646} $  & 0.76 \\
SSS120424:143145$-$355205  &  SSS1431$-$35 & 14:31:45 &$-$35:52:05  & 2018 Feb 15   &   300   &  HA   &  14.1  & $ 524  ^{+124 }_{-84 } $  & 0.16\\
CSS120617:152351+083606    &  CSS1523+08   & 15:23:51 &+08:36:06    & 2018 Apr 16   &   1200  &  D    &  19.0  & $ 2437 ^{+1275}_{-737} $  & 0.70\\
CSS120101:152731+181727    &  CSS1527+18   & 15:27:31 &+18:17:27    & 2018 Apr 16   &   1200  &  D    &  20.2  & $ 1850 ^{+1216}_{-684} $  & 1.27\\
CSS120527:153833$-$151719  &  CSS1538$-$15 & 15:38:33 &$-$15:17:19  & 2018 Apr 16   &   480   &  D    &  19.4  & $ 2461 ^{+2592}_{-1240}$  & 0.85\\
CSS160906:160346+193540    &  CSS1603+19   & 16:03:46 &+19:35:40    & 2018 Apr 16   &   480   &  P   &  19.6  & $ 261  ^{+15  }_{-14 } $  & 0.05\\
CSS120301:161823$-$102500  &  CSS1618$-$10 & 16:18:23 &$-$10:25:00  & 2018 Apr 17   &   1200  &  D    &  19.8  & $ 2698 ^{+2942}_{-1618}$  & 0.84\\
CSS160528:162350$-$121731  &  CSS1623$-$12 & 16:23:50 &$-$12:17:31  & 2018 Apr 16   &   1200  &  P    &  20.1  & $ 743  ^{+941 }_{-265 }$  & 0.28\\
CSS130421:172701+181421    &  CSS1727+18   & 17:27:01 &+18:14:21    & 2018 Apr 17   &   1200  &  D    &  21.2  & $ 774  ^{+1815}_{-371 }$  & 0.43\\
CSS150512:175608+265535    &  CSS1756+26   & 17:56:08 &+26:55:35    & 2018 Apr 17   &   480   &  D    &  18.5  & $ 1439 ^{+735 }_{-383 }$  & 0.30\\
CSS140629:201637$-$103060  &  CSS2016$-$10 & 20:16:37 &$-$10:30:60  & 2018 Apr 17   &   1200  &  HA   &  18.6  & $ 2575 ^{+2175}_{-1139}$  & 1.21\\
CSS150708:204247$-$095351  &  CSS2042$-$09 & 20:42:47 &$-$09:53:51  & 2018 Apr 16   &   1200  &  D    &  21.5  & $ 1591 ^{+1687}_{-702 }$  & 0.62\\
CSS121005:212625+201948    &  CSS2126+20   & 21:26:25 &+20:19:48    & 2018 Apr 17   &   480   &  D    &  18.7  & $ 1163 ^{+823 }_{-374 }$  & 0.38\\
MLS161006:214653$-$021820  &  MLS2146$-$02 & 21:46:53 &$-$02:18:20  & 2018 Apr 17   &   600   &  P    &  21.0  & $ 941  ^{+1133}_{-523 }$  & 0.70\\
\enddata                                                                                                         
\tablenotetext{a}{P: polar candidate, IP: IP candidate, D: disk system, HA: system with high-state accretion disk, E: extragalactic source, T: T Tauri star.  }
\tablenotetext{b}{MLS1050+10, MLS1137+00 and MLS1247$-$04 have negative parallaxes in the Gaia DR2, which usually indicates large distances. The values presented here are for reference only and are incompatible with the extragalactic classification.}
\end{deluxetable*}                                                      
\end{longrotatetable}                                                                                                                                                                                       

\newpage

\section{Observational characteristics of magnetic CVs}
\label{sec_classification_criteria}

\begin{figure*}
\epsscale{0.8}
\plotone{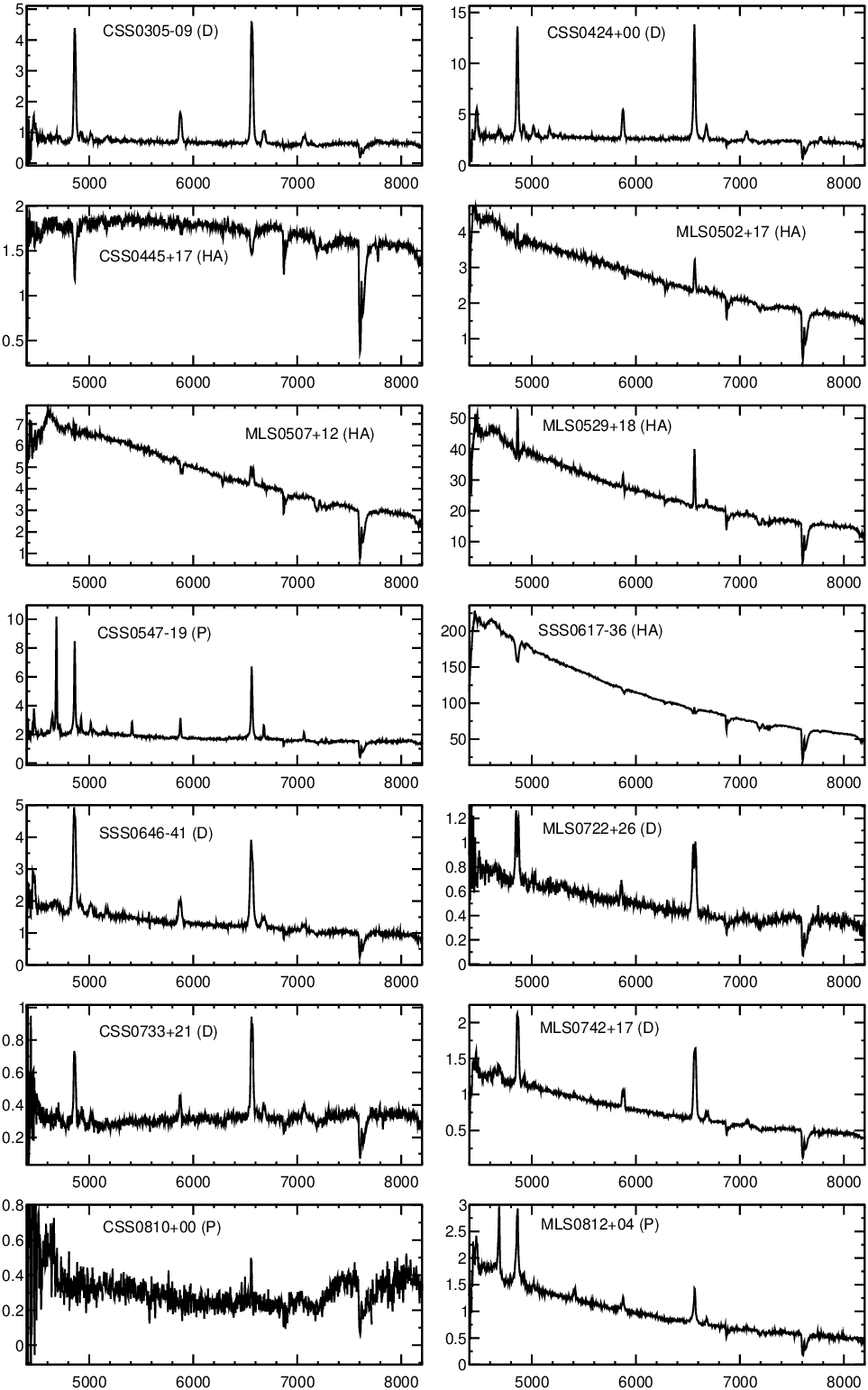}
\caption{Identification spectra arranged by right ascension. The vertical axes are F$_{\lambda}$ in units of $10^{-16}$ erg cm$^{-2}$ s$^{-1}$  \r{A}$^{-1}$ and the horizontal axes are Wavelengths, in \r{A}. \label{allspecs}
}
\end{figure*}

The spectra obtained in this survey are presented in Figure~\ref{allspecs}. We classified the sample in the following generic types: mCV candidates (polars and IPs), disk-dominated CVs (mainly dwarf novae -- DN -- in quiescence), CVs in high-accretion state (usually DN in outbursts or nova-like systems), T Tauri star, and extragalactic sources (active galactic nuclei - AGN).

AGN sources (discussed in Sec.~\ref{sec_extragal}) may present CRTS light curves that mimic the expected variation of polars, but their spectra are easily distinguishable from CV spectra because of their redshifted, broad emission lines. On the other hand, the distinction between non-magnetic CVs and mCVs may be much more challenging. 

Polar's light curves alternate irregularly between high and low states in time-scales of months or years, with amplitudes of 1 to 3 magnitudes. Unlike some IPs, they do not present outbursts. In shorter time-scales (tens of minutes to few hours) polars can show variations of brightness as large as $1-2$ magnitudes, associated to the WD spin synchronized with the orbital revolution, while IPs present smaller amplitudes (tenths of magnitudes) and may have shorter (minutes) periodic signals related to the unsynchronized WD spin.

Emission lines, mainly from hydrogen, are the most common spectral  indicator of mass accretion. In particular, magnetic CVs are usually identified by the presence of high-ionization lines of \ion{He}{2} and the \ion{C}{3}/\ion{N}{3} blend at 4650~\r{A}, which are originated in the accretion column. However, these features may be present in nova-like and dwarf novae spectra as well \citep{warner1995}. 
Due to the absence of accretion disks, the spectra of polars usually present narrow emission lines. 
Although low-inclination non-magnetic CVs can also display those narrow lines, in polars these are normally asymmetric since they are formed in different regions in the accretion flow as well as in the illuminated surface of the companion star. On the other hand, line profiles in IPs can be very similar to those of non-magnetic CVs, presenting large widths and even double peaks. 
In short, due to the lack of accretion disks, polars are more easily classified than IPs, which are observationally  more similar to non-magnetic CVs.

When in high accretion state, the spectra of polars have a flat or blue continuum and typically do not show conspicuous contribution from stellar components. When in low state, the optical spectrum of polars may show cyclotron humps or TiO bands from the cold stellar companion, and may be devoid of the \ion{He}{2} and the \ion{C}{3}/\ion{N}{3} emission features. Also in this state, polars may reveal the photosphere of the WD as Balmer absorptions as well as Zeeman features associated to the Balmer lines. 
Some objects in low accretion state may be accreting matter via magnetic siphoning of the wind of a secondary star that underfills its Roche lobe, in which case may be termed Pre-Polars \citep[PREPS;][]{2009A&A...500..867S,2015SSRv..191..111F}. Alternatively, ordinary polars in a prolonged low accretion state that is associated to Roche lobe overflow may be named Low Accretion Rate Polars \citep[LARPS;][]{2002ASPC..261..102S,2015SSRv..191..111F}. As both low accretion rate classes share several observational properties, the distinction between them is not straightforward. 

An empirical and quantitative criterion to identify mCVs spectroscopically was proposed by \citet{silber1986}:  H$\beta$ equivalent width (EW) is larger than 20~\r{A}, and \ion{He}{2} 4686\r{A} to H$\beta$ EW ratio
is larger than 0.4.
However, all these features should be understood as suggestive of magnetic accretion, but not as definitive conditions. 
Table \ref{fwhm} presents the FWHM and the equivalent width of the Balmer and \ion{He}{2} 4686~\r{A} lines of all objects classified as CVs in this work, including these mCV candidates. 
In Fig.~\ref{figLineRatio} we show a diagram with the equivalent widths of the \ion{He}{2} 4686~\r{A} versus H$\beta$ spectral lines for all CVs, observed in this work and in Paper I, for which the EW of H$\beta$ could be measured. Systems with the \ion{He}{2} 4686~\r{A} feature absent or too weak to be measured but with measurable H$\beta$ are represented in the diagram by data-points with ordinates equal to zero. The Silber's criteria for mCV classification are represented by straight lines in the diagram. The systems which we classified as mCVs by the spectra are mostly located inside the region of the Silber's criteria.
A discussion about this figure is presented in Section~\ref{sec_discussion}. 

Moreover, X-ray emission is not exclusive properties of mCVs, being observed in many non-magnetic CVs. Indeed, although a large fraction of mCVs were discovered by their X-ray emission, some mCVs are not detected in X-rays \citep[e.g.,][]{2015MNRAS.451.4183S,2014ASPC..490..389M}.

A conclusive classification as mCV often can only be achieved with follow-up observations. 

\section{CLASSIFICATION RESULTS}
\label{sec_classification}

\begin{figure*}
\epsscale{0.8}
\plotone{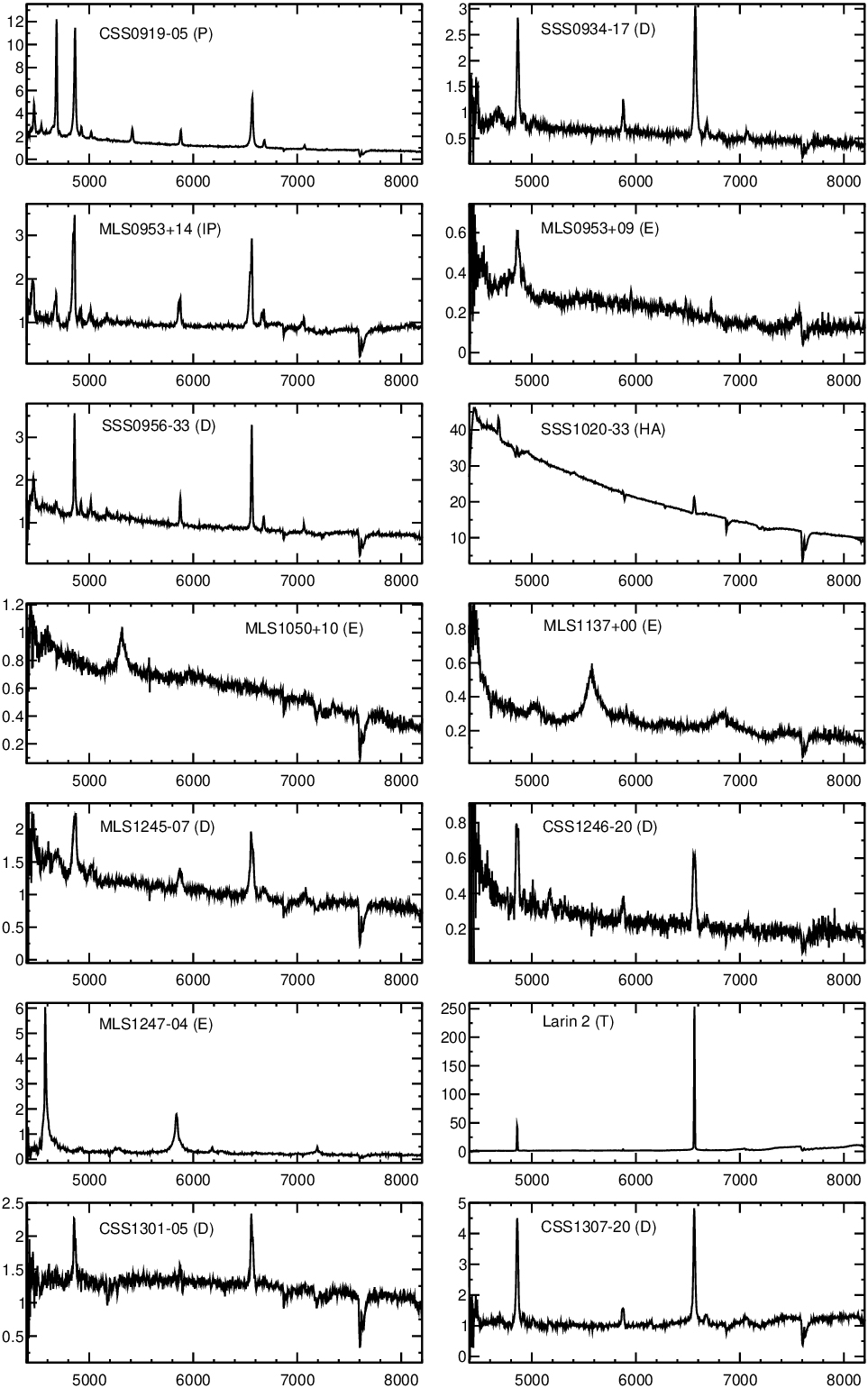}
\flushleft
    { Fig. 1.---\it{Continued}} 
\end{figure*}

\begin{figure*}
\epsscale{0.8}
\plotone{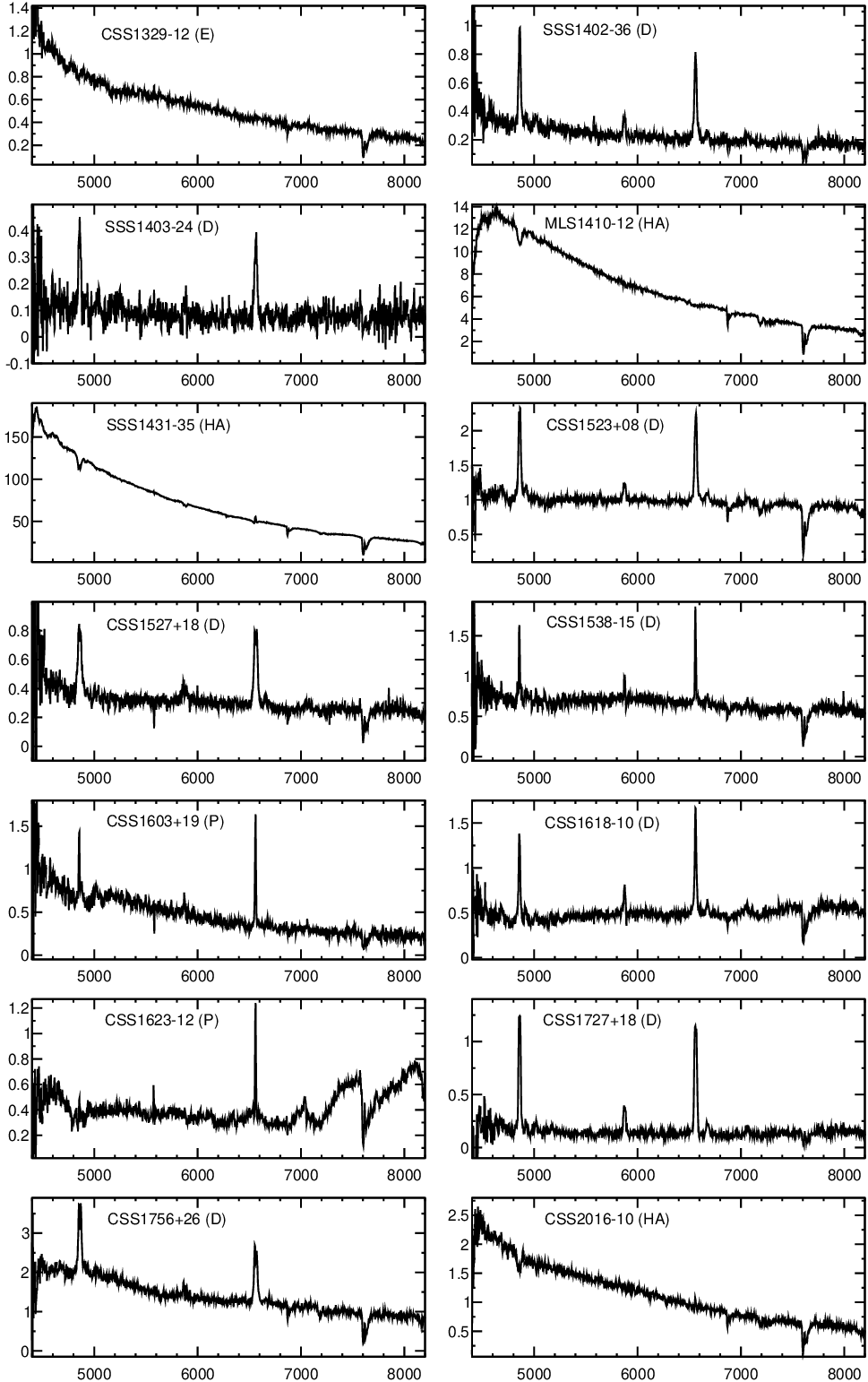}
\flushleft
    { Fig. 1.---\it{Continued}}
\end{figure*}

\begin{figure*}
\epsscale{0.8}
\plotone{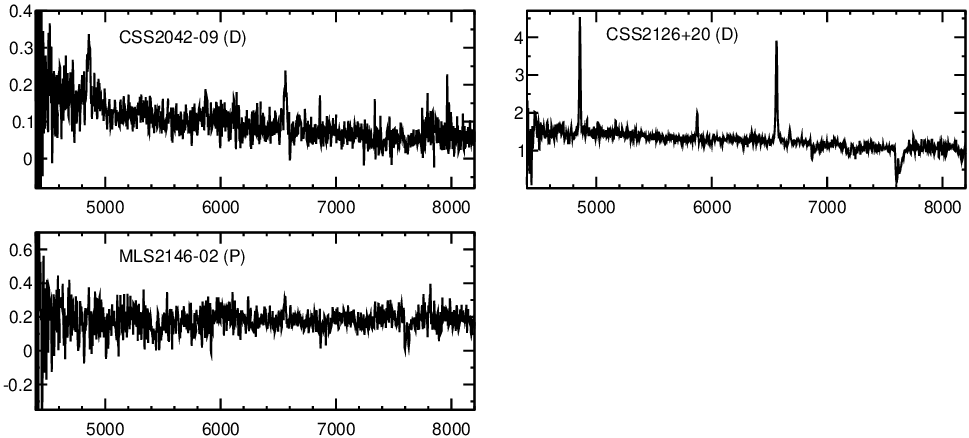}
\flushleft
    { Fig. 1.---\it{Continued}}
\end{figure*}


\begin{deluxetable*}{ccccccccc}
\tabletypesize{\scriptsize}
\tablecaption{Equivalent Width (EW) and Full Width at Half Maximum (FWHM\tablenotemark{\scriptsize{a}}) of \hheii, H$\beta$ and H$\alpha$ emission lines of the objects classified as CVs.  
\label{fwhm}}
\tablewidth{0pt}
\tablehead{
\colhead{Object} & \colhead{$-$EW He II } &\colhead{FWHM He II}  & \colhead{$-$EW H$\beta$ } &\colhead{FWHM H$\beta$ }  & \colhead{$-$EW H$\alpha$} &\colhead{FWHM H$\alpha$}  & \colhead{He II/H$\beta$} & \colhead{Type\tablenotemark{\scriptsize{b}}}\\
\colhead{} & \colhead{({\AA}) } &  \colhead{(km s$^{-1}$)}&\colhead{({\AA})} &  \colhead{(km s$^{-1}$)}& \colhead{({\AA}) }  &  \colhead{(km s$^{-1}$) }& \colhead{} & 
}
\startdata
CSS0305$-$09  &   7   & 1600     &     134      &    1400    &            197    &    1170     &     0.05     &  D        \\
CSS0424+00    &   13  & 2300     &     94       &    1180    &            113    &    1020     &     0.14     &  D         \\
CSS0445+17    &   \nodata &  \nodata     &     -10      &    1330    &            -5     &    1500     &     \nodata    &  HA     \\
MLS0502+17    &   p\tablenotemark{\scriptsize{c}}   &  p       &     2        &    550     &            7      &    710   &   \nodata   &  HA  \\
MLS0507+12    &   \nodata &  \nodata     &     1        &    \nodata     &            6      &    1680     &     \nodata      &  HA \\
MLS0529+18    &   \nodata &  \nodata     &     5        &    660     &            12     &    610      &     \nodata & HA   \\
CSS0547$-$19  &   55  &  650     &     50       &    730     &            55     &    650      &     1.10   &  P  \\
SSS0617$-$36  &   p   &  p       &     -5       &    2130    &            2      &    500      &     \nodata    &  HA    \\
SSS0646$-$41  &   p   &  p       &     70       &    1830    &            80     &    1500     &     \nodata   &  D    \\
MLS0722+26    &   \nodata &  \nodata     &     35       &    2600    &            66     &    1960     &     \nodata    &  D              \\
CSS0733+21    &   15  &  1100    &     41       &    1420    &            64     &    1190     &     0.37   &  D          \\
MLS0742+17    &   10   &  1800    &     23       &    1520    &            52     &    1390     &     0.44    &  D        \\
CSS0810+00    &   \nodata &  \nodata     &     \nodata      &    \nodata     &            25     &    510      &     \nodata &  P    \\
MLS0812+04    &   25  &  870     &     26       &    1250    &            24     &    850      &     0.96    &  P      \\
CSS0919$-$05  &   85  &  800     &     110       &    1000    &            115    &    630      &     0.77    &  P     \\
SSS0934$-$17  &   20:  &  2300    &     75       &    1260    &            153    &    1130     &     0.27    &  D        \\
MLS0953+14    &   35  &  1660    &     85       &    1670    &            65     &    1480     &     0.41     &  IP         \\
SSS0956$-$33  &   7   &  2100    &     35       &    770     &            48     &    600      &     0.20     &  D             \\
SSS1020$-$33  &   4   &  1330    &     1:        &    870     &            6      &    790      &     4.00:    &  HA                \\
MLS1245$-$07  &   15  &  4190    &     34       &    2540    &            41     &    1800     &     0.44      &  D            \\
CSS1246$-$20  &   \nodata &  \nodata     &     57       &    1970    &            90     &    1680     &     \nodata   &  D    \\
CSS1301$-$05  &   p   &  p       &     20       &    1380    &            29     &    1270     &     \nodata     &  D         \\
CSS1307$-$20  &   p   &  p       &     94       &    1300    &            90     &    900      &     \nodata     &  D              \\
SSS1402$-$36  &   \nodata &  \nodata     &     78       &    1440    &            95     &    1160     &     \nodata   &  D    \\
SSS1403$-$24  &   \nodata &  \nodata     &     110       &    1740    &            225    &    1470     &     \nodata   &  D  \\
MLS1410$-$12  &   \nodata &  \nodata     &     -4       &    2410    &            \nodata    &    \nodata      &     \nodata     &  HA     \\
SSS1431$-$35  &   p   &  p       &     -5       &    2280    &            3      &    530      &     \nodata        &  HA           \\
CSS1523+08    &   15   &  3400    &     50      &    1600    &            40     &    1220     &     0.30     &  D        \\
CSS1527+18    &   \nodata &  \nodata     &     70       &    2740    &            85     &    1870     &     \nodata      &  D            \\
CSS1538$-$15  &   5   &  330     &     27       &    660     &            26     &    480      &     0.19      &  D       \\
CSS1603+19    &   \nodata &  \nodata     &     22       &    550     &            53     &    470      &     \nodata       &  P     \\
CSS1618$-$10  &   17      &  2200        &     65       &    1070    &            50     &    790      &     0.26       &  D        \\
CSS1623$-$12  &   \nodata &  \nodata     &     15:       &    500:     &            29     &    400      &     \nodata     &  P     \\
CSS1727+18    &   \nodata &  \nodata     &     270      &    1380    &            318    &    1300     &     \nodata      &  D      \\
CSS1756+26    &   \nodata &  \nodata     &     30       &    2310    &            48     &    1740     &     \nodata      &  D      \\
CSS2016$-$10  &   \nodata &  \nodata     &     -7       &    1970    &            4      &    560      &     \nodata      &  HA     \\
CSS2042$-$09  &   \nodata &  \nodata     &     65       &    2240    &            46     &    1510     &     \nodata      &  D      \\
CSS2126+20    &   \nodata &  \nodata     &     31       &    730     &            38     &    620      &     \nodata      &  D      \\
MLS2146$-$02  &   p       &  p           &     p        &   p        &            32     &    1030     &     \nodata      &  P      \\
\enddata
\tablenotetext{a}{The FWHM were measured from Voigt profile fiting. }
\tablenotetext{b}{The object types are described in Table~\ref{obs}}
\tablenotetext{c}{The symbol p means that the line is present in the spectrum but its S/N is low and no measurements were possible. }
\end{deluxetable*}

\subsection{Magnetic CV Candidates}
\label{sec_polars}

This section comprises seven polar candidates and one IP candidate.

\begin{figure}
 \includegraphics[width=\columnwidth]{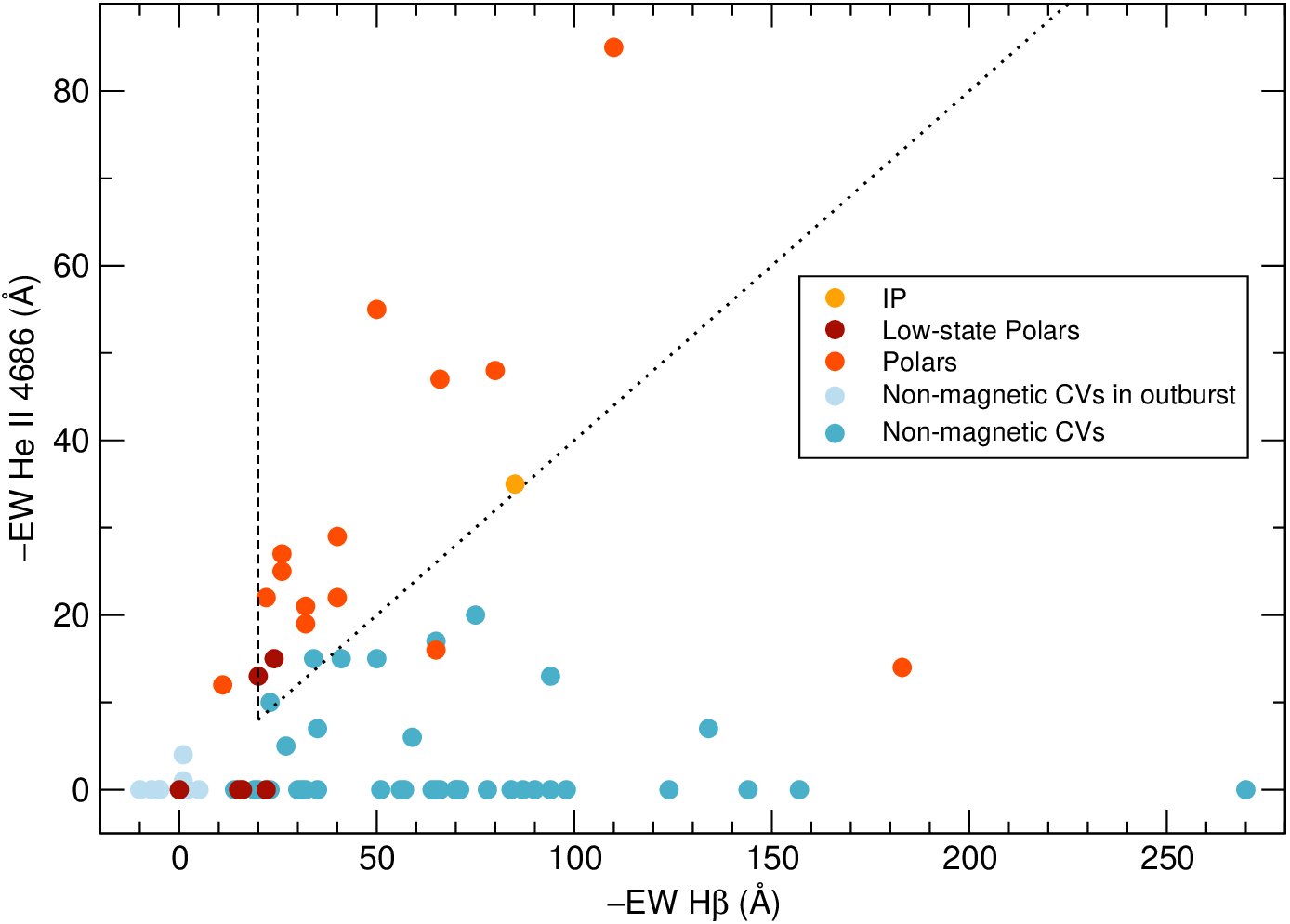}
  \caption{Equivalent widths (EW) of the \ion{He}{2} 4686~\r{A} versus H$\beta$ emission lines, for the magnetic and non-magnetic CVs presented in this work and in Paper I. The lines delimit the locus of mCVs from the Silber's criteria: $-$EW~H$\beta$~$>$~20~\r{A} (dashed line) and EW~\ion{He}{2}/EW~H$\beta$~$>$~0.4 (dotted line). Objects having the \ion{He}{2} spectral line absent or too weak to be measured are represented with $-$EW~\ion{He}{2}~=~0.
  }
\label{figLineRatio}
\end{figure}

\subsubsection{CSS0547-19 (CSS101108:054711-192525)}

This object was detected as a transient source in the CSS survey on 2010 November 8. Its CRTS light curve is highly variable and ranges from $\sim$ 18 to 20 mag in timescales of days. ROSAT detected a possible X-ray counterpart of CSS0547-19 in the ROSAT All-Sky Survey (RXS J054711.6-192518), located 10.7 arcsec from the optical source's coordinates, and hence within the nominal X-ray positional uncertainty of 12 arcsec. 
The Gaia DR2 distance to CSS0547-19 is $868^{+474}_{-241}$ pc (Table~\ref{obs}).
The identification spectrum is dominated by narrow and asymmetric emission lines of the Balmer series and \heii over a slightly blue continuum, besides lines of \ion{He}{1}, \ion{He}{2}~5412~\r{A}, \ion{Fe}{2} and the Bowen \ion{C}{3}/\ion{N}{3} complex at $\sim 4640$~\r{A}. The EW ratio of the \heii to H$\beta$ lines is 1.1. These observational features are typical of polar CVs. From the continuum flux at $\sim 5500$~\r{A} of our SOAR spectrum, we estimated a magnitude of $\sim18.3$, indicating that the source was observed at its high CRTS brightness state.

\subsubsection{CSS0810+00 (CSS100108:081031+002429 = CRTS CSS100108 J081031+002429)}

This transient has an extremely variable light curve in the CRTS, fluctuating from 21.5 to 17 mag in short timescales. It is not detected as an X-ray source. From photometric monitoring, \citet{2012MNRAS.421.2414W} classify CSS0810+00 as a polar, for the similarity to the light curve of V348 Cen, and determine a period of P=0.080660 d. That r-band light curve was obtained in high state, in which the system varies from around 17.5 to 19.5 mag.
\citet{2014MNRAS.441.1186D} present an spectrum with a flat continuum and Balmer, HeI, and HeII emission lines, which 
they describe as a typical spectrum of a quiescent CV.

\begin{figure}
 \includegraphics[width=\columnwidth]{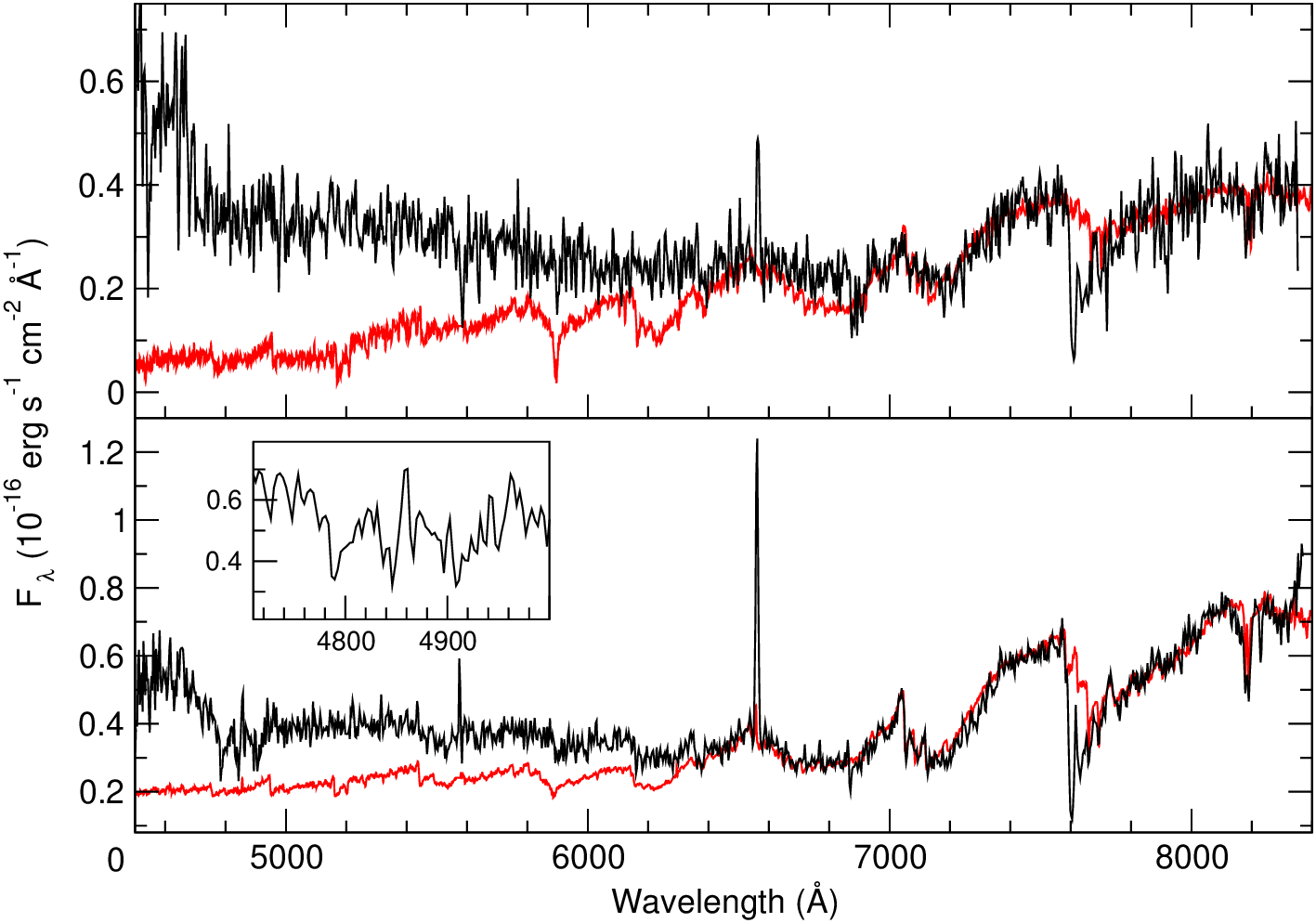}
  \caption{Spectra of the low-accretion state polars CSS0810+00 and CSS1623-12. Top panel: Spectra of CSS0810+00 (black line) and M2 SDSS template (red line). Bottom panel: CSS1623-12 (black line) and M5 SDSS template (red line). The inset graphic shows the H$\beta$ Zeeman components in the CSS1623-12 minus M5 residual spectrum.}
\label{figresiduos}
\end{figure}

Our SOAR spectrum  was obtained at $V \sim 20.3$~mag, i.e., while in low state. It shows strong TiO features of a cool star, narrow H$\alpha$ emission and a broad hump at $\sim4600$~\r{A}. We compared the red continuum and the TiO features with SDSS spectral templates of dwarf M stars from \citet{2007AJ....133..531B} and found that the best match is a M$2\pm1$ spectral type (Fig.~\ref{figresiduos}).
The expected spectral type from the orbital period-mass relation from \cite{2011ApJS..194...28K} is M4.6. The difference is within the spread of observations around this relation. 
The spectrum of CSS0810+00 is typical of a polar in low accretion state.  Despite being noisier, its spectrum is very similar to the spectrum of CSS1623-12 (Sec. \ref{sec_1623}), although the low S/N prevents the detection of the narrow absorptions near H$\beta$ seen in CSS1623-12.

\subsubsection{MLS0812+04 (MLS170128:081210+040352 = CRTS CSS091017 J081210+040352)}

MLS0812+04 was discovered as a variable source by the MLS survey in 2017, and was afterwards identified as the known transient CSS080228:081210+040352. The CRTS light curve spans $\sim 4000$ days and shows rapid (days) variability between 17.5 and 19.5 mag, with some excursions to 20 - 20.5 mag. 
\citet{2014MNRAS.441.1186D} suggest it is an eclipsing system.
No X-ray or EUV counterpart is found within $10^3$ arcsec from its coordinates. \citet{2018AJ....156...58B} list a distance of  $1356^{+543}_{-324}$ pc to MLS0812+04. Our spectrum presents a blue continuum and strong \heii and Balmer lines in emission, as well as \ion{He}{1}, \ion{He}{2}~5412~\r{A}, and weak Bowen complex. Its \ion{He}{2}/H$\beta$ ratio is 0.96. The flux of the SOAR spectrum corresponds to 18.8 mag.
We classify this object as a polar CV.

\subsubsection{CSS0919-05 (CSS110114:091937-055519 = SSS110606:091937-055519 = CRTS J091936.6-055519)}

The CRTS light curve of CSS0919-05 shows alternating brightness states between 16 and 20 mag, with superimposed $\sim 1$ mag rapid variability. 
An X-ray transient (OT~J091936.6-055519)
at 35~arcsec from the position of CSS0919-05 was detected by Swift XRT PC. 
The Swift XRT has a nominal PSF of 18~arcsec at 1.5~keV, so it is not completely discarded that the transient is related to the optical source.
There is also a possible Galex counterpart under identification 2736405182818550631, with FUV=19.77 mag and NUV=18.92 mag. The Gaia Data Release 2 (2018) has no recorded parallax for this object. This object was selected as a QSO candidate by \citet{2007ApJ...664...53A}, with a probability of being a QSO of $P_{QSO}=0.73$.  Nonetheless, the spectrum obtained at SOAR is consistent with a Galactic object (no redshift) and is typical of polar CVs, presenting high intensity \heii and Balmer emission lines, with \ion{He}{2}/H$\beta$ = 0.77, and less intense \ion{He}{1} emission feature  and \ion{C}{3}/\ion{N}{3} Bowen complex. 
\citet{2014MNRAS.441.1186D} presents a spectrum of this object very similar to ours. From the estimated magnitude of 18.6 mag, CSS0919-05 was nearly at CRTS low state.

\subsubsection{MLS0953+14 (MLS110303:095308+145837 = CRTS J095308.2+145837)}

This system was discovered and classified as a polar candidate (1RXS~J095308.6+145841) by a selection of variable X-ray sources with soft X-ray spectrum from the ROSAT All Sky Survey
\citep{1995ASPC...85...99B,1999A&A...347...47B}. 
However, \citet{1995ASPC...85...99B} note that these selection criteria might also uncover IPs with soft X-ray emission. In addition, they suggest a period of 102 min for this system. 
The optical light curve used in this period determination and  optical data obtained subsequently are presented by \citet{2008AIPC..984...32K}. The light curve is typical of magnetic accretion onto a WD. These authors also call attention to a possible decrease in period from 6316~s (105.27~min in 1995) to 6245~s (104.08~min in 2002), which should be confirmed by additional data.
MLS0953+14 was not detected in radio by
the Jansky VLA survey \citep{2017AJ....154..252B}, while IR WISE data phased with the 102 min period presented an amplitude of $\Delta$W1 = 0.5 mag \citep{2015ApJS..219...32H}. From IR colors, these authors suggest the presence of cyclotron emission associated to a magnetic field with $B\leq 30$ MG. The $\sim10$~yr long CRTS light curve of MLS0953+14 shows large variability between 17 and 20 mag. 
This range of variability is consistent with the magnitude variation along the period from \citet{2008AIPC..984...32K}.
\citet{2008NewA...13..133A} estimated a distance of 364 pc to this object using the Period-Luminosity-Colors (PLC) method, while the Gaia distance is $448^{+61}_{-48}$ pc \citep{2018AJ....156...58B}, with a fractional parallax uncertainty of $f=0.12$. 
\citet{2019arXiv191108338H} have recently presented the spectrum of this object and classify it as a polar candidate. The spectrum shows a marginally detected \ion{He}{2} emission line. 

Our identification spectrum of MLS0953+14 was obtained at V = 19.0 mag. It
presents intense Balmer lines in emission, as well as \ion{He}{1} and \heii emission features. All these features have similar line profiles, being asymmetrical with extended blue wings and showing evident double peaks. 
The H$\beta$ EW is $-85$~\r{A} and the \ion{He}{2}/H$\beta$ ratio is 0.41, which meet the Silber's empirical criteria for mCV classification. No clear signs of the \ion{C}{3}/\ion{N}{3} Bowen complex or of the \ion{He}{2}~5412~\r{A} line are seen in this spectrum. 
The object's spectrum resembles that of EX~Hya \citep{1987MNRAS.228..463H}.
In the light of all these spectral characteristics, particularly the double-peaked lines, which point to the presence of an accretion disc, we classify MLS0953+14 as an IP. 
In this case, the 
photometric period \citep{2008AIPC..984...32K} should probably be associated to the WD rotation, and not to the orbital period.

\subsubsection{CSS1603+19 (CSS160906:160346+193540)}

The CRTS circulars classify CSS1603+19 as a CV, likely of the polar type, from its light curve. It alternates between high brightness states, ranging from 16.5 to 18 mag, and low states, from 18.6 to 20 mag, in timescales of tens of days. \citet{2017RNAAS...1a..29T} report an orbital period of $P_{orb}=1.37$ h, determined from unpublished radial-velocity spectroscopy. The Gaia distance to this system is only 261 pc \citep{2018AJ....156...58B}, yet no X-ray source is detected close to its coordinates. The SOAR spectrum was obtained at 19.6 mag (therefore at low brightness state) and displays very narrow emission lines of H$\alpha$ and H$\beta$, besides weak \ion{He}{1}.
Absorption troughs, which might be associated to Zeeman splitting components, are seen around H$\beta$. No clear sign of a cool secondary star is present in the spectrum. Based on the light curve and on the spectral features, we classify CSS1603+19 as a polar candidate, observed in low accretion state. More observations of this object could settle this classification.

\subsubsection{CSS1623-12 (CSS160528:162350-121731)}
\label{sec_1623}

This source presents a CRTS light curve with well-defined high (at average $\sim17$ mag) and low (at $\sim19.5$ mag) brightness states, with some occasional measurements in the transition between those states. In addition, at any state it shows variability with about 1 mag amplitude in a timescale of days. A possible X-ray counterpart - 1RXS J162349.0-121707 - lies 31 arcsec from its coordinates, compared with a positional uncertainty of 25 arcsec. The SOAR identification spectrum was obtained at low state, with V=20.1 mag, and is similar to the spectrum of CSS0810+00 described above. It is dominated by TiO features of a cool star and quite narrow H$\alpha$ and H$\beta$ emission lines, and also presents the \ion{Na}{1} absorption doublet at 8183, 8194~\r{A} from the secondary. The continuum is nearly flat, with a broad hump at 4600~\r{A} and narrow absorption features around the Balmer lines associated to Zeeman components. It is quite similar to the spectrum of AM Her in low state \citep{1981ApJ...243L.157S}. By fitting the red portion of the spectrum with template dwarf M SDSS spectra from \citet{2007AJ....133..531B}, we determined a M5$\pm1$ spectral type for the companion star (Fig.~\ref{figresiduos}). 
This indicates that the system is probably below the period gap.
The spectrum suggests the presence of a cyclotron hump centered at around 4600\AA.

We visually compared the H$\beta$ Zeeman components, in the difference spectrum, with SDSS magnetic WDs template spectra \citep{2009A&A...506.1341K} and estimated a magnetic field B between 8 and 12 MG for the WD. If this value for B is correct, the first and strongest cyclotron humps should be in the mid to near-infrared spectral regions, which is inconsistent with a cyclotron origin for the broad hump near 4600~\r{A}.
We classify this object as a polar candidate in low accretion state. More observations are needed to constrain the WD magnetic field value.

\subsubsection{MLS2146-02 (MLS161006:214653-021820)}

MLS2146-02 has an extremely variable CRTS light curve, ranging from 16.7 to 20.5 mag in timescales of few weeks. An X-ray counterpart, \mbox{1RXS~J214653.0-021812}, is located 8 arcsec from its coordinates. Because of these features, it was suggested to be a polar CV, possibly eclipsing, in the CRTS circulars. The Gaia distance to this object is 941 pc \citep{2018AJ....156...58B}. Unfortunately our exploratory spectrum caught MLS2146-02 in an extremely faint state, at V=21 mag, so it is dominated by noise. Even so, one can see weak H$\alpha$ line in emission, and also emission features that could be associated to H$\beta$ and \hheii. Spectra obtained at a higher brightness state is still necessary to verify if it is really a polar type CV as indicated by its optical light curve and X-ray emission.

\subsection{Non-magnetic CVs}

In this section we describe CVs that present spectral characteristics associated to quiescent accretion disks or objects that do not have strong evidence of magnetic accretion. Those objects present prominent Balmer lines in emission, often double-peaked in higher inclination systems, together with weaker emissions of \ion{He}{1}, \ion{He}{2} and heavier elements, over a continuum with a blue slope. These spectral features are mainly associated to dwarf novae in quiescence, but may also be found in the (few known) short period IPs like, for instance, SDSS J2333 \citep[= V598 Peg --][]{2005AJ....129.2386S,2007MNRAS.378..635S}. 
More observations of objects in this section, however, could reveal that some objects of them harbor a magnetic WD.
Systems with disks in high accretion states will be described in the next section.

\subsubsection{CSS0305-09 (CSS101014:030535-092005)}
\label{sec_CSS0305-09}

CRTS classifies this blue transient as a polar candidate because of the detection of a possible X-ray counterpart by ROSAT (1WGA J0305.5-0919) in addition to its high level of optical variability. However, there is another object with V$\sim16$~mag at an angular distance of 20 arcsec from the CSS0305-09 coordinates, which is also inside the ROSAT error circle. The CRTS light curve varies between 18 to 20 mag, with a single measurement at $\sim16.5$ mag in October 2014 that could be associated to an DN outburst, being back to 19.9 mag in the next measurement taken one month later. Our identification spectrum corresponds to V=19.3 mag, consistent with the median level of the CRTS light curve. It shows single-peaked Balmer lines, \ion{He}{1}, and \ion{Fe}{2} all in emission, besides weak \hheii. The continuum is nearly flat, with a very small slope to the blue.

\subsubsection{CSS0424+00 (CSS081030:042434+001419)}

The CRTS light curve of CSS0424+00 displays a base level around 18 mag with $\sim 1$ mag dispersion. In addition, it shows outbursts reaching 16 mag and fadings to magnitude $\sim 20$, which last few months. No X-ray source is detected close to its coordinates. The identification spectrum, obtained at 17.9 mag, is typical of disk systems, with strong Balmer and weak \ion{He}{1} and \ion{Fe}{2} emission lines. Broad and weak \heii emission is also present, with a EW ratio of 0.14 relative to H$\beta$.

\subsubsection{SSS0646-41 (SSS130329:064610-410416)}

SSS0646-41 has a possible X-ray counterpart detected by the ROSAT All Sky Survey (\mbox{1RXS~J064611.7-410406}) and also by the XMM, in both cases slightly outside the error circles.
The CRTS light curve varies between 17 and 18 mag, with several measurements reaching 14.5 to 16 mag.
Interestingly, the baseline flux of this object in CRTS data seems to vary smoothly in long term scale within 1-mag amplitude.
Our spectrum of this object was obtained at $V=18.6$ mag, therefore at its lowest registered CRTS brightness. It presents broad and asymmetric Balmer and \ion{He}{1} emission lines over a blue continuum, while a weak \heii feature is also visible. By using an automated script based on the CRTS light curves, \citet{2016MNRAS.456.4441C} classified this source as a DN.

\subsubsection{MLS0722+26 (MLS150118:072211+260255 = CRTS CSS090128 J072211+260255)}

This object is classified as a CV by \citet{2014MNRAS.441.1186D}. Its CRTS light curve shows multiple outbursts at $17-18$ mag, a base level at $19-20$ mag, and possible eclipses reaching $21.5$ mag. It has no X-ray counterpart. The SOAR spectrum (at $V=19.6$ mag) presents double-peaked Balmer lines, consistent with an eclipsing disk system.

\subsubsection{CSS0733+21 (CSS091111:073339+212201 = CRTS CSS091111 J073339+212201)}

CSS0733+21 is a CRTS CV candidate \citep{2014MNRAS.441.1186D} with X-ray (1RXS~J073340.7+212208) and GALEX-DR5 \citep{2011Ap&SS.335..161B} counterparts.
Its light curve varies between 19 and 20 mag, with sporadic measurements at $16-17$ mag, while the identification spectrum has a flat continuum and Balmer lines, besides \ion{He}{1} and weak \hheii.

\subsubsection{MLS0742+17 (MLS110309:074223+172807 = CRTS MLS110309 J074223+172807)}

This object has a highly variable CRTS light curve around $19-20$ mag with numerous outbursts at \mbox{$\sim18$~mag} and likely eclipses reaching 21 mag. The SOAR spectrum (V=19 mag) presents double-peaked  Balmer and \ion{He}{1} lines, in addition to broad and weak \heii line, over a blue continuum.

\subsubsection{SSS0934-17 (SSS100505:093417-174421)}

This CRTS transient was discovered and classified as a variable star in May 2010. The CRTS light curve shows long-term (months) up and down transitions, between 17 and 20 mag, which are unlike DN outbursts. Our SOAR spectrum was obtained in a faint state ($V=19.1$ mag) and presents the usual features of a disk accretor system, as strong Balmer and weak \ion{He}{1} emission lines. It also exhibits broad \heii with EW ratio of $\sim0.27$ relative to H$\beta$. No X-ray counterpart is detected at its position.

\subsubsection{SSS0956-33 (SSS120111:095652-331216 = CRTS SSS120111 J095652-331216)}

This CRTS transient is a CV candidate in \citet{2014MNRAS.441.1186D}. The light curve from CRTS spans 6.5 years and is consistent with a bimodal brightness distribution (high/low states) or with frequent outbursts. The quiescent level lies at $\sim 19$ mag, with a superimposed 1 mag amplitude variation, while the high-brightness measurements reach 16 mag. The identification spectrum was obtained in quiescent state and presents narrow Balmer features with symmetric profiles, as well as narrow \ion{He}{1} and \heii lines, over a blue continuum.

\subsubsection{MLS1245-07 (MLS150511:124539-073706)}

This object was discovered by CRTS in May 2015 and suggested to be a polar type CV. Its light curve presents a quiescent level with long-term (years), low-amplitude ($19-19.5$ mag) variability and one outburst in 2015 (when it was detected by CRTS) reaching 14.5 mag. Our exploratory spectrum displays broad H$\alpha$ and H$\beta$ with asymmetric profile, possible associated to a double peak, besides broad \ion{He}{1} and \heii lines.

\subsubsection{CSS1246-20 (CSS120222:124602-202302 = CRTS J124602.0-202302)}

\citet{2014MNRAS.441.1186D} classify this system as a CV. The CRTS light curve varied between 17.5 and 19 mag from 2005 to 2012, when it started to show outbursts reaching 15.5 mag and drops down to 19.8 mag. No X-ray counterpart is known for this object. The SOAR spectrum (at V = 20.4~mag) presents broad double-peaked lines of Hydrogen and \ion{He}{1}, but no hint of \hheii.
\citet{2010MNRAS.402..436W} show a spectrum presenting the same emission lines, but with a quite steeper blue continuum.

\subsubsection{CSS1301-05 (CSS140430:130136-052938)}

The CRTS light curve of CSS1301-05 shows a quiescent level at 18.5 mag, with $\sim 1$ mag amplitude variability, and half a dozen outbursts as bright as $16-17$~mag. The SOAR spectrum has a flat continuum with double-peaked Balmer lines in emission. \heii is marginally detected. Also clearly visible in the spectrum are absorptions identified as the Mg~b triplet at 5167, 5172, 5183~\r{A}, the Ca+Fe feature at 5269~\r{A} and the \ion{Na}{1} D doublet, indicative of a late G or early K spectral type secondary. This indicates that the orbital period of CSS1301-05 should be long, possibly above $7-8$ h.

\subsubsection{CSS1307-20 (CSS120304:130755-202628 = CRTS CSS120304 J130755-202628)}

CSS1307-20 is classified as a CV by \citet{2014MNRAS.441.1186D}. The CRTS light curve has a baseline level at $18-19$ mag and few measurements at $14.5-16$ mag. Our spectrum, obtained at $\sim 19$ mag, is typical of a DN is quiescent state, with narrow Balmer emissions, \ion{He}{1}, besides marginally detected \hheii.

\subsubsection{SSS1402-36 (SSS130313:140204-365251)}

This transient, discovered by CRTS in 2013, shows numerous outbursts at $15.5 - 17$ mag over a quite scattered quiescent state at 18 to 20 mag, possibly related to eclipses. In the SOAR spectrum obtained at 20.4~mag, however, the Balmer and \ion{He}{1} emission lines are single-peaked, with no sign of the \heii feature.

\subsubsection{SSS1403-24 (SSS130312:140311-240135)}

This object was classified by the CRTS as a CV candidate because of its light curve, which presents fast variability ranging from 21 to 17.3 mag. Our identification spectrum confirms this classification, showing Balmer and \ion{He}{1} emission lines, over a flat continuum.

\subsubsection{CSS1523+08 (CSS120617:152351+083606)}

\citet{2014MNRAS.441.1186D} classify CSS1523+08 as a CV candidate. The CRTS light curve is extremely variable and bimodal, with a scattered quiescent level ranging from 20.5 to 18.1 mag 
and very frequent measurements in the range from 17 to 18 mag. The distance to this source, as measured from Gaia DR2 parallax, is 2437 pc \citep{2018AJ....156...58B}. It has an X-ray counterpart (2XMM J152351.2+083605) detected by XMM-Newton. The SOAR exploratory spectrum shows H$\beta$ more intense than H$\alpha$ and EW ratio of \heii to H$\beta$ of 0.30, besides weaker emission lines of \ion{He}{1} and a flat continuum. The H$\beta$ and \ion{He}{1} profiles presents double peaks which, however, are not seen in H$\alpha$. This spectrum was obtained in the quiescent level, with an estimated magnitude of V=19.0~mag.

\subsubsection{CSS1527+18 (CSS120101:152731+181727 = CRTS CSS120101 J152731+181727)}

CSS1527+18 is another CV candidate from \citet{2014MNRAS.441.1186D}. Its light curve has fast variability between 19 and 21 mag, likely associated to eclipses, and some outbursts reaching 17 mag. In one outburst, detected in 2017 December 8 and described by Gaia Photometric Science Alerts\footnote{ESA Gaia, DPAC and the Photometric Science Alerts Team (\url{http://gsaweb.ast.cam.ac.uk/alerts})} (Gaia17dfn), it brightened by 2 mag in less than 2 days. Our spectrum, obtained in quiescence at 20.2 mag, displays broad, double-peaked Balmer and \ion{He}{1} emission lines, supporting the eclipse assumption.

\subsubsection{CSS1538-15 (CSS120527:153833-151719)}

The CRTS light curve of CSS1538-15 displays a baseline level at 19 mag, with $\sim1.5$ mag variation amplitude, and few outbursts between 16.4 and 17.3 mag. Gaia also detected one outburst in 2018 August 8 (Gaia18cbw). \citet{2014MNRAS.441.1186D} classified this source as a CV candidate. The identification spectrum, observed at an estimated magnitude of V=19.4 mag, has a flat continuum and narrow lines of Hydrogen and Helium in emission. The EW ratio of \heii to H$\beta$ is 0.19.

\subsubsection{CSS1618-10 (CSS120301:161823-102500 = CRTS CSS120301 J161823-102500)}

This object has a CRTS light curve varying from 19 to 20 mag and very frequent outbursts spanning from 18.5 to 17 mag. The exploratory spectrum, obtained in quiescent brightness state, presents the usual Balmer lines in emission and weak and broad Helium features, including \hheii, besides \ion{Na}{1} D in absorption. This is object is also cited a CV candidate by \citet{2014MNRAS.441.1186D}.

\subsubsection{CSS1727+18 (CSS130421:172701+181421)}

This transient was discovered independently, in 2013 April, by CRTS and by ASAS-SN \citep[\mbox{ASASSN-13ag} --][]{2013ATel.5052....1S}. This line of sight matches some high energy sources: GALEX~J172700.7+181420, 1RXS~J172709.0+181234, and  XMMSL2~J172700.3+181422. 
\citet{2012A&A...548A..99W} classify this X-ray source as a QSO. In the CRTS light curve, it is a very faint source, with measurements below $19.5 - 20$ mag, and one recorded outburst reaching 15 mag, when it was discovered as a transient. Our spectrum, corresponding to V = 21.5~mag, shows double-peaked Balmer and \ion{He}{1} emission lines associated to an accretion disk. The X-ray object seems to be a persistent source, which is at odds with the faint optical source.

\subsubsection{CSS1756+26 (CSS150512:175608+265535)}

The CRTS light curve of CSS1756+26 is quite variable between 17.2 and 20 mag. Our spectrum, obtained at $\sim18.5$ mag, shows double-peaked Balmer and \ion{He}{1} superimposed on a blue continuum.
In spite of the absence of outbursts, we could not find any evidence of magnetic accretion in the system.

\subsubsection{CSS2042-09 (CSS150708:204247-095351)}

This source is highly variable in the CRTS light curve, with points scattered between 21 and 17 mag. The SOAR spectrum, obtained at 21.5 mag, is noisy. Still, it clearly presents Balmer lines in emission, with H$\beta$ stronger than H$\alpha$, as well as \ion{He}{1}. The spectrum quality allows us to confirm the object as CV, but it is not possible to verify if it is a magnetic system.

\subsubsection{CSS2126+20 (CSS121005:212625+201948 = SDSS J212625.08+201946.3)}

CSS2126+20 is a CV discovered as a CRTS transient by \citet{2014ApJS..213....9D}. Its CRTS light curve is quite peculiar, with measurements between 16 to 17 mag and also fast changes to a state between 18.5 to 19.5 mag. \citet{2016JBAA..126..178S} performed a photometric monitoring campaign and discovered superoutbursts with a period of 66.9 d and short outbursts separated by $\sim 11$ d, besides superhump with a period of 0.088 d. Therefore they classify this object as a SU UMa DN with one of the shortest known superoutbursts periods. Our SOAR spectrum was obtained at quiescent state (V $\sim 18.7$ mag) and reveals
intense and narrow Balmer lines as well as weak \ion{He}{1} and \ion{Fe}{2}.

\subsection{Systems with Disks in High Accretion State}

Objects in this section have as characteristic spectral features a steep blue continuum and Balmer lines (specially H$\beta$) in broad absorption, frequently with a narrow emission line core. These characteristics are usually associated to dwarf novae in outbursts or nova-like CVs, where the increased continuum and the broad absorption are due to a thick, high accretion-rate disk. Since the Balmer decrement is steeper in emission lines than in absorption lines, the higher series members are usually seen in absorption while H$\alpha$ remains in emission. Alternatively, these spectral features could be related to quiescent dwarf novae at extremely low mass-transfer rates, like in GW Lib \citep{2000AJ....119..365S}. In the latter case, the broad absorptions would arise from the white dwarf photosphere since the accretion disk is not the dominant source of light in the system. The analysis of long-term light curves like the ones provided by the CRTS is a way to discriminate between these scenarios, since no outbursts are expected in nova-like systems. Besides, information on the brightness state of the object by the time of the spectrum observation allows to distinguish the outbursting dwarf nova from quiescent, low mass-transfer rate dwarf nova.

\subsubsection{CSS0445+17 (CSS140903:044529+173745)}

CSS04456+17 presents an extremely variable light curve in CRTS. It virtually displays two brightness states, at $V\sim17.5$ mag and at $V\sim19.5$~mag, with many fast transitions between both. The exploratory spectrum, obtained in the high brightness state, shows a moderately flat continuum and no emission lines. Among the absorption lines we identify H$\alpha$ (with a perceptible central emission), H$\beta$, and narrow features of \ion{He}{1} 4922 \r{A}, \ion{He}{2} 5016 \r{A}, \ion{Fe}{2} 5169 and 5317 \r{A} and \ion{O}{1} 7772 \r{A}. Also present are the interstellar absorption lines of \ion{Na}{1} 5890 and 5896 \r{A} and the diffuse interstellar band (DIB) at 6284 \r{A}. Despite the nearly flat continuum, these characteristics point to an erupting DN classification.

\subsubsection{MLS0502+17 (MLS101203:050253+171041 = CRTS J050253.1+171041)}

The CRTS light curve of MLS0502+17 ranges from a 18.0-19.5 mag base level to less frequent measurements reaching 16.5 mag. Our identification spectrum was obtained at $\sim 17.7$ mag. This source has no detected X-ray counterpart. The identification spectrum is characteristic of a dwarf nova in outburst, evolving to its maximum, with narrow Balmer emission features over a blue continuum.  As in other spectra of DN in the same phase of the outbursts, the H$\beta$ emission is superimposed on a broad absorption trough \citep[e.g. SS Cyg,][]{1984ApJ...286..747H}. The interstellar absorption lines of \ion{Na}{1} D and the DIB at 6284 \r{A} are also seen.

\subsubsection{MLS0507+12 (MLS121018:050716+125314 = CRTS J050716.2+125314)}

This transient has a light curve varying from 21 to 19 mag and several outbursts reaching up to 16 magnitude. \citet{2013PASJ...65...23K} detected superhumps in its light curve, classifying it as a SU UMa-type DN. Our spectrum was obtained in eruption and shows Balmer lines in emission with a deep central absorption, besides absorptions of \ion{Na}{1} D and the DIB at 6284 \r{A}, all over a blue continuum, attesting the DN classification.

\subsubsection{MLS0529+18    (MLS101214:052959+184810 = CRTS MLS101214 J052959+184810) }

MLS0529+18 was detected as a transient by the CRTS survey in 2010, but has been known as a variable star since 1935 \citep[HV~6907, NSV~02026 -][]{1935BHarO.901...20H,2002IBVS.5298....1W}. Outbursts observed by CRTS led to the classification as a dwarf nova. It is probably associated to the X-ray counterpart 1RXS~J052954.9+184817, which is at a separation of 0.98 arcmin. \citet{2016PASJ...68...65K} classify MLS0529+18 as a SU UMa dwarf nova by the occurrence of superoutbursts in 2015 and 2016. The CRTS light curve shows measurements varying from 18.0 to 13.5 mag. Our spectrum was obtained in the rise to -- or fading from -- an outburst, at $\sim 15.2$ mag, and is quite similar to the spectrum of MLS0502+17, with H$\beta$ in narrow emission in the center of a shallow absorption feature, H$\alpha$ in emission, and a continuum with a blue slope. This object was observed by the LAMOST survey in quiescent state \citep{2019arXiv191108338H}.

\subsubsection{SSS0617-36    (SSS120320:061754-362654 = CRTS J061753.9-362655)     }

SSS0617-36 was discovered by CRTS in 2010. Its light curve ranges from a low-level state at $\sim 18.5$ - 17 mag to repeated outbursts reaching 14 - 15 mag. \citet{2012MNRAS.421.2414W} performed photometric time series of this source in 2010, during quiescence, and obtained non-sinusoidal, large-amplitude light curves with a 0.143 d periodicity. Their phased diagram shows an unusual shape for a dwarf nova CV. We estimate that our spectrum was obtained at 13.6 mag, therefore near maximum outburst brightness. This is in agreement with the observed spectral features: a steep blue continuum with H$\beta$ in absorption and H$\alpha$ with a short central emission over a broad absorption, characteristic of a hot, optically thick disk. No X-ray sources are detected within 15 arcmin from SSS0617-36.

\subsubsection{SSS1020-33    (SSS120215:102042-335002 = CRTS SSS120215 J102042-335002)     }

The CRTS light curve of this object is quite variable, showing frequent outbursts that reach 14.5 mag and a base level ranging from 18 to 17 mag. Our spectroscopic observation caught it in high state, at 15.4 mag. This spectrum shows the usual features of a DN in outburst, like the blue continuum associated to a hot disk and the development of broad absorption for the higher series of the Balmer lines, besides a strong \heii emission.

\subsubsection{MLS1410-12    (MLS130513:141002-124809 = CSS080502:141002-124809 = CRTS J141002.2-124809)}

MLS1410$-$12 was suggested to be a CV candidate by \citet{2009ApJ...696..870D}. The CRTS light curve shows measurements at $\sim 20$ mag and also at 16-17~mag. It is a SDSS source, but no spectrum was acquired by that survey. Our identification spectrum was obtained at $\sim16.6$~mag. It is characteristic of a DN in outburst near maximum brightness, with increased blue continuum and H$\beta$ in broad absorption. No sign of H$\alpha$ is seen, neither in emission nor in absorption.

\subsubsection{SSS1431-35 (SSS120424:143145-355205 = CRTS J143145.3-355205)}

SSS1431-35 is a cataclysmic variable candidate from \citet{2014MNRAS.441.1186D}. It is a bright source whose CRTS light curve varies from a fluctuating 18 mag base level to outbursts at $\sim13$ mag. There is a highly probable X-ray counterpart detected at about 5 arcsec from the SSS1431-35 coordinates, by XMM (XMMSL2 J143145.4-355202) with a count rate of 0.73 counts sec$^{-1}$ and by ROSAT PSPC (2RXP J143144.9-355210) with 0.12 counts sec$^{-1}$. We took the SOAR spectrum at V$\sim14.1$ mag. We classify this system as a DN at outburst by its spectrum and brightness state.

\subsubsection{CSS2016-10    (CSS140629:201637-103100)} 

The CRTS light curve of this system is highly variable, showing a bimodal brightness distribution around 17--18 and 19--21 ~mag. It is a distant source (d = $2575^{+2175}_{-1139}$) by Gaia parallax measurements \citep{2018AJ....156...58B}. We observed it at $V=18.6$ mag and the spectrum shows the typical features of a DN in outburst, with H$\beta$ in absorption and weak H$\alpha$ in emission over a blue continuum.

\subsection{Variable star of the T Tauri type}
\label{larin}
\subsubsection{Larin 2 (SSS J124850.8-412654 = EQ J124850-412654)}

\begin{figure}
 \includegraphics[width=\columnwidth]{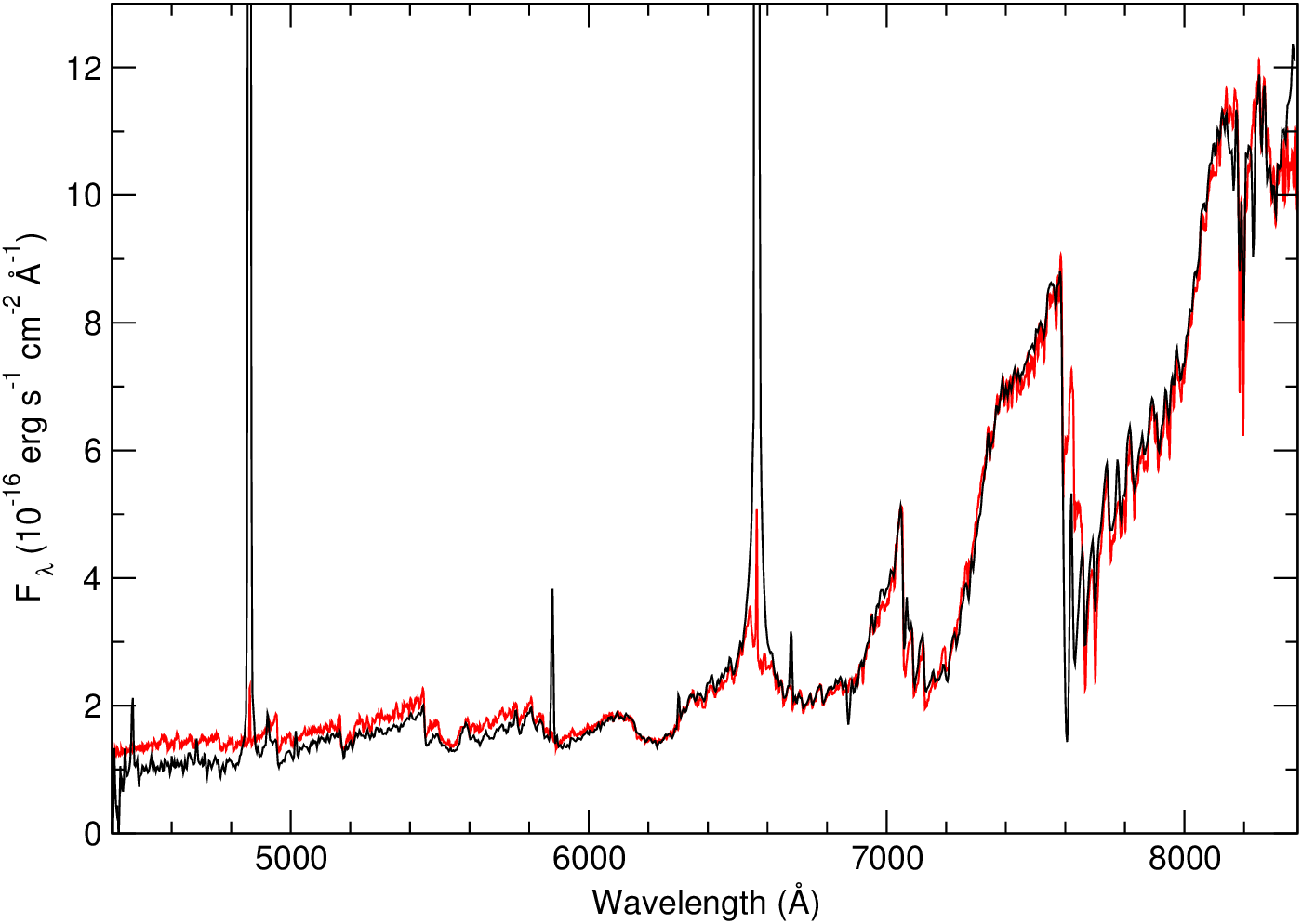}
  \caption{Average spectrum of Larin 2 (black line) and fitted dwarf M7 SDSS template spectrum (red line) from \citet{2007AJ....133..531B}. The ordinates are adjusted to emphasize the low-intensity features.   }
\label{figlarin}
\end{figure}

This object was discovered by I. Larin  on March 2018 as a large amplitude transient \citep{2018ATel11401....1L}. It presents peculiar colors ($FUV-W1=8.3$) and an X-ray flux 
that varies by a factor of 12 in XMM-Newton and Chandra telescopes. The CRTS light curve of this transient varies between 17 and 18 mag, with one point at 15.8 mag. \citet{2018ATel11401....1L} also describe photometric time series performed at the 0.5 m Chilescope Observatory, which shows 
fast variability in the range 15.9 to 16.5 mag, and suggest a polar classification. \citet{2018arXiv180704574D} report a 2 mag outburst of Larin 2 in ASAS-SN and its similarity to the pre-polar MQ Dra. We obtained 15 spectra of Larin 2 at SOAR in two consecutive nights on April 2018. The spectra are dominated by narrow and extremely intense Balmer emission, in addition to TiO features,  \ion{He}{1} and hints of [\ion{Ne}{2}] 5755~\r{A} and [\ion{O}{1}] 6300~\r{A} (Fig.~\ref{figlarin}). 
The TiO features are well fitted by a dwarf M7 spectral type template from \citet{2007AJ....133..531B}.
The radial velocity curves obtained from the Balmer emission lines in those spectra span $\sim2$ h and are flat, with an amplitude lower than 18~km~s$^{-1}$ and 3.6~km~s$^{-1}$ r.m.s. 
We obtained polarimetric data covering $\sim2$~h at the 1.6m P\&E telescope at OPD/LNA in Brazil on 2018 March 18 and 19, but no circular polarization was detected.
These observational characteristics, together with the spectral features, X-ray and optical variability, and infrared excess, indicate that Larin 2 is a classic T Tauri (CTT) star like, e.g., BP Tau or HH31 IRS2. \citet{2019arXiv190713152P} obtained a VLT/X-shooter spectrum of Larin2 and concluded that it is likely an young stellar object, in accordance with our results.

\subsection{Extragalactic Objects}
\label{sec_extragal}

Transients presented in this section were selected as mCV candidates due to their CRTS light curves, which show variability that could be interpreted as distinct high/low states. However, The SOAR identification spectra have shown that they are active galactic nuclei (AGN), which frequently share observational features with CVs, like long-term photometric variability and colors. The main spectral features leading to the classification as extragalactic sources are the redshifted  \ion{Mg}{2} 2798 \r{A} or Ly$\alpha$ emission lines. To the best of our knowledge, these are the first classification and redshift measurements for these sources. All of them, except MLS0953+09 and CSS1329-12, have negative Gaia parallaxes, which are the result of noisy Gaia measurements and imply that the distance is likely to be very large \citep{2015PASP..127..994B}.

\subsubsection{MLS0953+09 (MLS120223:095322+094127)}

The light curve of this CRTS CV candidate shows a single outburst, sampled by 5 data points, that varied from magnitude 19 to magnitude 14 in a timescale of only one day in 2005 March 11. Since then, the source remained in the range from 21 to 19 mag. The SOAR spectrum, obtained at $V\sim 20.4$ mag, presents a dominant broad emission feature at $4861$ \r{A} that could be identified as H$\beta$, but shows no sign of H$\alpha$. Instead, half a dozen weak and narrow emission lines are seen in the spectrum. We therefore identify the strong line as \ion{Mg}{2} 2798~\r{A}, while the weak lines are clearly associated to [\ion{Ne}{5}] 3347-3427~\r{A}, [\ion{Fe}{7}] 3587~\r{A}, [\ion{O}{2}] 3728~\r{A}, [\ion{Ne}{3}] 3870~\r{A}, H$\delta$, and H$\gamma$, among others. These lead to the classification of MLS0953+09 as an AGN at redshift 0.738. The Gaia distance of 604 pc to this transient (with a large fractional parallax uncertainty of $f=0.62$) is incompatible with that classification.

\subsubsection{MLS1050+10 (MLS110226:105051+102134)}

The CRTS photometric data of MLS1050+10 presents fast (one day) $\sim0.5$ mag variation superimposed on a bimodal light curve with 22 mag and 20.5 mag average values, besides one large and fast outburst that reached 19.5 mag. Our exploratory spectrum was obtained at high photometric level, at V=19.4 mag. It presents a conspicuous broad emission line which we identify as \ion{Mg}{2} 2798~\r{A}, therefore leading to a classification as a Type I AGN at redshift $z=0.90$, further confirmed by the detection of weak emission lines of [\ion{Ne}{5}] 3427 \r{A}, [\ion{O}{2}]~3728 \r{A}, and [\ion{Ne}{3}]~3870 \r{A}.

\subsubsection{MLS1137+00 (MLS121203:113751+004218)}

This object was suggested as a likely CV in the CRTS circulars. Its light curve presented a slow (years) transition from a low state, with average 21 mag and 1.5 mag scatter, to an also variable V$\sim20$ mag high state. \citet{2016AJ....152....5D} included it in a sample of 15 CVs with Kepler K2 monitoring, but did not discuss this object due to its low brightness. No X-ray source is known close to its coordinates. The SOAR spectrum is consistent with a Type I AGN at $z=0.99$ based on the presence of the broad \ion{Mg}{2} 2798~\r{A} emission line and several blended UV \ion{Fe}{2} multiplets.

\subsubsection{MLS1247-04 (MLS140518:124709-040758)}

MLS1247-04 is a blue SDSS source that shows fast brightness variability in CRTS data, changing by almost 3 magnitudes in timescales of weeks. For that reason it was classified as a CV candidate in the CRTS circulars. Its light curve is quite variable, ranging from about 22 to 18 mag. The SOAR spectrum, however, clearly presents emission features of Ly $\alpha$, \ion{O}{1} 1304~\r{A}, \ion{Si}{2} 1307~\r{A}, \ion{C}{4} 1549~\r{A}, and \ion{He}{2} 1640~\r{A}, among numerous other lines, which classify this source as an AGN at redshift $z=2.765$.

\subsubsection{CSS1329-12 (CSS130419:132918-121622)}

This object was selected for its CRTS light curve, which looks bimodal with measurements around 19-20~mag and around 16-17.5~mag. Our spectrum was obtained in the lower brightness state, at 19.5 mag. It presents a steep blue continuum, but is almost featureless, except for the telluric absorptions near 6800 and 7600~\r{A}. Its CRTS light curve variation and the featureless blue spectrum are consistent with a BL Lac classification. The fractional parallax uncertainty for this object is very large, $f=3.04$, which could be associated to a relatively distant source. The absence of clear spectral features hinders the determination of its redshift.

\section{SUMMARY AND DISCUSSION}
\label{sec_discussion}

In this paper we present the results of the second part of our spectroscopic survey to search for mCVs, from targets selected form the CRTS catalog. In Paper I, from a sample of 45 targets we found 32 CVs, 22 of which classified as mCV candidates, being 13 out of those 22 reported as mCV for the first time. Now, from a sample of 45 targets to which we obtained SOAR spectra, we found 39 CVs, 5 AGNs and 1 T Tauri star. 
Among the 39 objects we classify as CVs, 8 show characteristic spectra of mCVs (7 polars and 1 IP), while 9 CVs 
present features typical of high accretion state systems. The remaining 22 objects do not present spectral features that allow us to unambiguously classify them as mCVs.
From the group of  8 mCV candidates,  6 are new classifications (CSS0547-19, MLS0812+04, CSS0919-05, CSS1603+19, CSS1623-12, and MLS2146-02) and 2 were already published as mCVs (CSS0810+00 and MLS0953+14). MLS0953+14 has been suggested to be a polar candidate for its X-ray behavior but our spectrum leads to a classification as IP, and CSS0810+00 has been previously classified as a polar system from the shape of its optical light curve. 

Three out of the seven polar candidates from this work are not detected as X-ray sources, which again shows (as also noted in Paper I) that X-ray detection is not a necessary condition for the classification as a polar (or as a mCV, see \citealt{2014ASPC..490..389M}).

Three polars (CSS0810+00, CSS1603+19 and CSS1623-12) were observed in low accretion state, as indicated by their spectral features and also by their brightness when compared to the CRTS light curves. CSS0810+00 and CSS1623-12 present spectra very similar to the spectrum of AM Her in low state, with TiO absorption features from a cold secondary star and narrow Balmer emissions. The companion's photospheric features allowed the estimation of their spectral type. Zeeman absorption components around H$\beta$, present in the spectrum of CSS1623-12, yield an estimate of B~$\sim 8-12$ MG for the WD magnetic field. The lower signal-to-noise ratio in the spectrum of CSS0810+00 hindered the detection of H$\beta$ Zeeman components. CSS1603+19 does not present clear signs of the secondary TiO features, but show evidence of H$\beta$ Zeeman components.

In Fig.~\ref{figLineRatio}, seven objects (from Paper I and also from this work) that we classify as mCVs fall out of the locus of the Silber's criteria for mCV classification. Four of them are low-accretion state polars (CSS0810+00, CSS1503-22, CSS1603+19 and CSS1623-12), for which the \ion{He}{2} 4686\r{A} is absent. Among the remaining three outlier polars (1RXS1002-19, CSS1127-05 and SSS1944-42, see Paper I), two could be considered peculiar in some way: 1RXS1002-19, which has a FWHM of 2\,000~km~s$^{-1}$, very high for a polar, and CSS1127-05, which has a large equivalent width value for H$\alpha$ (EW = 204~\AA).
In the same diagram, two non-magnetic CVs (MLS0742+17 and MLS1245-07) are found inside the mCV region. Both present broad \ion{He}{2} 4686\r{A} emission feature, double-peaked Balmer emission lines and do not present long-term light curves typical of DNe. These objects deserve more observations to a better understanding of their nature. 

We classify 22 
objects as CVs with quiescent disks. The majority of the spectra show the usual Balmer and \ion{He}{1} emission lines (often double-peaked) characteristic of quiescent DNs. These classifications, however, do not rule out the possible presence of IPs among these objects, since systems like the short-period IPs SDSSJ~2333 \citep{2005AJ....129.2386S,2007MNRAS.378..635S} and HT~Cam \citep{1998A&A...335..227T}, for instance, may present spectra that are undistinguishable from the spectra of quiescent DNs.
The remaining CVs in our sample are systems with disks at high accretion state. By comparing the time of our spectral observations with their CRTS light curves we suggest that all of them are dwarf novae in eruption.

Our main goal in this work is to improve the statistics of mCVs. The classification of a target as a mCV is not trivial, since usually no unique observational technique can provide a definite classification. The strategy of spectroscopic snapshot observations of targets selected by variability criteria is a first step in the direction of our goal. Almost 90 percent of the targets in the sample presented in this paper were CVs, while about 15 percent turned out to be magnetic. Multi-technique, time-resolved follow-up observations of these mCV candidates will be used to verify the proposed classifications and to unravel their detailed nature.

\vspace{5mm}

\acknowledgments

This paper is based on observations obtained at the Southern Astrophysical Research (SOAR) telescope, which is a joint project of the Minist\'{e}rio da Ci\^{e}ncia, Tecnologia, Inova\c{c}\~{o}es e Comunica\c{c}\~{o}es (MCTIC) do Brasil, the U.S. National Optical Astronomy Observatory (NOAO), the University of North Carolina at Chapel Hill (UNC), and Michigan State University (MSU), and on observations obtained at the Observatório do Pico dos Dias/LNA. 

ASO acknowledges São Paulo Research Foundation (FAPESP) for financial support under grant \#2017/20309-7. CVR would like to thank the financial support under grants \#2013/26258-4 (São Paulo Research Foundation, FAPESP) and \#303444/2018-5 (CNPq). MSP thanks CAPES for financial support under grant \#88887.153742/2017-00. IJL thanks FAPESP by grants \#2018/05420-1, \#2015/24383-7, \#2013/26258-4.

The CSS survey is funded by the National Aeronautics and Space Administration under Grant No. NNG05GF22G issued through the Science Mission Directorate Near-Earth Objects Observations Program.  The CRTS survey is supported by the U.S. National Science Foundation under grants AST-0909182 and AST-1313422.

This work has made use of data from the European Space Agency (ESA) mission \href{https://www.cosmos.esa.int/gaia}{\it Gaia}, processed by the {\it Gaia} Data Processing and Analysis Consortium (\href{https://www.cosmos.esa.int/web/gaia/dpac/consortium}{DPAC}). Funding for the DPAC has been provided by national institutions, in particular the institutions participating in the {\it Gaia} Multilateral Agreement.

This research has made use of data, software and/or web tools obtained from the High Energy Astrophysics Science Archive Research Center (HEASARC), a service of the Astrophysics Science Division at NASA/GSFC and of the Smithsonian Astrophysical Observatory's High Energy Astrophysics Division, and of the VizieR catalogue access tool, CDS, Strasbourg, France (DOI: 10.26093/cds/vizier). The original description of the VizieR service was published in A\&AS 143, 23.

\vspace{5mm}

%

\facility{SOAR (Goodman HTS)}


\software{IRAF \citep{1986SPIE..627..733T}}









\end{document}